\documentclass{JHEP3}
\usepackage{epsfig}
\usepackage{graphicx}%
\def\be{\begin{equation}}
\def\ee{\end{equation}}
\def\bea{\begin{eqnarray}}
\def\eea{\end{eqnarray}}
 
\def\lesssim{\mathrel{\hbox{\rlap{\hbox{\lower4pt\hbox{$\sim$}}}\hbox{$<$}}}}
\def\gtrsim{\mathrel{\hbox{\rlap{\hbox{\lower4pt\hbox{$\sim$}}}\hbox{$>$}}}}

\title{Cosmology With Negative Potentials}

\author{Gary Felder\\
CITA, University of
Toronto, 60 St. George Street, Toronto, ON M5S 3H8, Canada\\ E-mail: \email{felder@cita.utoronto.ca}}
\author{Andrei Frolov\\
    CITA, University of
Toronto, 60 St. George Street, Toronto, ON M5S 3H8, Canada\\ E-mail: \email{frolov@cita.utoronto.ca}}
\author{Lev Kofman\\
    CITA, University of
Toronto, 60 St. George Street, Toronto, ON M5S 3H8, Canada\\ E-mail: \email{kofman@cita.utoronto.ca}}
\author{Andrei Linde\\
    Department of Physics, Stanford University, Stanford, CA 94305,
USA\\
    E-mail: \email{linde@physics.stanford.edu}, ~~ http://physics.stanford.edu/linde}
\received{\today}        
 \preprint{CITA-2002-04\\ SU-ITP-02/05\\ \hepth{0202017}\\ February 4, 2002}

\abstract{ We investigate cosmological evolution in models where the effective potential $V(\phi)$ may become negative for some values of the field $\phi$.  Phase portraits of such theories in space of variables $(\phi,\dot\phi,H)$ have several qualitatively new features as compared with phase portraits in the theories with $V(\phi) > 0$. Cosmological evolution in models with potentials with a ``stable'' minimum at $V(\phi)<0$ is similar in some respects to the evolution in models with
potentials unbounded from below.  Instead of reaching an AdS regime
dominated by the negative vacuum energy, the universe reaches a turning point
where its energy density vanishes, and then it contracts to a singularity
with properties that are practically independent of $V(\phi)$.  We apply our
methods to investigation of the recently proposed cyclic universe scenario.
We show that in addition to the singularity problem there are other problems that need to be resolved in order to realize a cyclic regime in this scenario. We propose several modifications of this scenario and conclude that the best way to improve it is to add a usual stage of inflation after the singularity and use that
inflationary stage to generate perturbations in the standard way.}

\keywords{eld.pbr.ctg.sgm}

\begin{document}

\section{Introduction}

Since the invention of inflationary cosmology
\cite{Starobinsky:te}-\cite{Hybrid}, the theory of the evolution of
scalar fields in an expanding universe has been investigated quite
extensively, both at the classical and the quantum level. While many
features of scalar field cosmology are well understood, the overall
picture remains somewhat incomplete.  In this paper we will extend the
investigation of scalar field cosmology to models with negative
effective potentials. We are also going to bring together several
other issues, such as the impact of radiation and particle production
on the onset of inflation. This will allow us to get a better
understanding of various possibilities that may appear in scalar field
cosmology.  

We are going to use a general approach based on the investigation of
3d phase portraits that show the behavior of the scalar field $\phi$,
its velocity $\dot\phi$, and the Hubble constant $H = {\dot a\over
a}$. We will see that the phase portraits of models with $V(\phi) > 0$
and with $V(\phi) < 0$ are qualitatively different and that additional
changes appear when one adds matter and/or radiation.

There are several reasons to study cosmology with negative potentials.
The first one is related to the cosmological constant problem.  The
simplest potential used in inflationary cosmology is $V(\phi)= { 1
\over 2} m^2\phi^2$ \cite{Chaot}.  One can add to this potential a
small cosmological constant $V_0$ without changing any features of
inflation. A small positive $V_0 \sim 10^{-120}$ (in Planck units)
would be sufficient to describe the present acceleration of the
universe in a de Sitter-like state. But why should $V_0$ be so small
and positive? What would happen for $V_0 <0$? Does the
post-inflationary universe with $V_0 <0$ behave like anti-de Sitter
space, which is so popular in M-theory?

Rather unexpectedly, the answer to this question appears to be
negative: After a long stage of inflation the universe with $V_0<0$
cannot approach an AdS regime; instead of that it collapses
\cite{Banks:1995dt,N8,Fastroll}. In this paper we will study cosmological behavior
in a large class of theories with negative potentials and explain why
the universe in these theories stops expanding and eventually
collapses.
 
Another reason to study theories with negative potentials is provided
by the investigation of cosmology in N=2,4,8 gauged
supergravity. Recently it was found that in all known versions of
these theories potentials with extrema at $V(\phi)>0$ are unbounded
from below. Despite this fact, such models can, under certain
conditions, describe the present stage of acceleration of the universe
\cite{N8,Fastroll}.

One more reason is related to a formal connection with warp
factor/bulk scalar dynamics in brane cosmology.  It has recently
been shown that the equations for the warp factor and scalar field in
brane cosmology with a scalar field potential $V(\phi)$ are similar to
the equations for the scale factor and scalar field in 4D cosmology
with the opposite potential $-V(\phi)$ \cite{Felder:2001da}.  This
reveals an interesting relation of cosmology with negative potentials and warp
geometry with positive potentials.

Finally, cosmology with a negative potential $V(\phi)$ is the basis of
the recently proposed ``cyclic universe'' model \cite{Cyclic} based in
part on the ekpyrotic scenario \cite{KOST}. However, unlike in the
ekpyrotic scenario \cite{KOST}, the authors of \cite{Cyclic} assume,
in accordance with \cite{pyrotech}, that the scalar field potential
$V(\phi)$ at large $\phi$ is positive and nearly constant. As a
result, the universe experiences ``superluminal expansion''
(inflation) that helps to solve some of the cosmological
problems. In this sense  cyclic scenario, unlike the ekpyrotic scenario of Ref. \cite{KOST}, is a specific version of inflationary theory rather than an alternative to inflation.\footnote{  One should note, however, that this is a very
specific kind of inflation that is possible only if the universe is
exponentially large all the time. Thus the large size of the universe
is not explained by inflation in this model but rather required for
it.}  Then the scalar field rolls to a minimum of its effective
potential with $V(\phi)<0$, the universe contracts to a singularity,
re-emerges and again enters a stage of inflation. This scenario
inherits many unsolved problems of the ekpyrotic model
\cite{pyrotech}, including the singularity problem \cite{Seiberg}. 
The authors assume that the universe can pass through the singularity
and that one can use perturbation theory and specific matching
conditions at the singularity to calculate density perturbations in
the post-big bang universe generated by processes prior to the
singularity \cite{Khoury:2001zk}.  This issue is rather controversial
\cite{Lyth:2001pf}.  The possibility of achieving a cyclic regime
depends on various assumptions concerning the creation of
matter and the acceleration of the scalar field during the big bang.

The idea that the big bang is not the beginning of the universe but a
point of a phase transition is quite interesting, see
e.g. \cite{Tolman-1931}-\cite{PBB}. However, the more assumptions
about the singularity one needs to make, the less trustworthy are the
conclusions. In this respect, inflationary theory provides us with a
unique possibility to construct a theory largely independent of any
assumptions about the initial singularity. According to this theory,
the structure of the observable part of the universe is determined by
processes at the last stages of inflation, at densities much smaller
than the Planck density. As a result, observational data practically
do not depend on the unknown initial conditions in the early universe.

Since the cyclic scenario does
require repeated periods of inflation anyway, it would be nice to
avoid the vulnerability of this scenario with respect to the unknown
physics at the singularity by placing the stage of inflation before
the stage of large scale structure formation rather than after it.

In order to achieve this goal we will examine the conditions that are
necessary for the existence of the cyclic regime in the model of
Ref. \cite{Cyclic} and then check whether the model can be modified in
a way that would not require various speculations about the behavior
of matter, the scalar field, and density perturbations near the
singularity.

Our paper will thus consist of two parts. The first part will contain
a general study of scalar field cosmology with positive and negative
potentials. The second part will be devoted to a  more speculative subject, it will include application of our general results to the cyclic scenario.  

In Section \ref{archetypes} we will describe several basic regimes
that are possible in scalar field cosmology: the universe can be
dominated by potential energy, by kinetic energy, by the energy
density of an oscillating scalar field, or by matter or radiation. The
discussion of these four distinct regimes will help us to understand
the phase portraits of the universe that we are going to draw in the
subsequent sections.

Section \ref{portraits} will describe the use of phase portraits for
studying cosmological evolution. We will write the evolution equations
for the field and scale factor in the form of three first order
equations plus one time dependent constraint.  The
solutions to these equations can then be represented as trajectories
in phase space, clearly showing the possible ways the universe can
evolve in different situations. Finally, by using a Poincar\'{e}
projection we can map the entire phase space onto a finite sphere,
thus allowing the complete set of possible trajectories to be easily
seen.

In Section \ref{positive} we will apply these methods to models with
positive definite potentials. Such potentials have been extensively
studied before  with the use of phase portraits
\cite{Khalat,KLS}. We study them here partly to introduce the methods
we are using and to provide a point of comparison for the negative
potentials of the following section. We also present some new results
 concerning the effects of matter and radiation on the
development of inflation.

In Section \ref{negative} we show the phase portraits for a model
where the effective potential can become negative. We discuss general
properties of such models, and in particular the ways in which they
differ from the models of the previous section.  One of our major
conclusions is that such models generically enter a stage of
contraction. In Section \ref{switch} we will examine in detail the
transition from expansion to contraction in models of this type.

Many of the features of scalar field cosmology that we are going to
discuss are model-independent. The phase portraits in Sections
\ref{positive}-\ref{switch} all use the simplest model $V(\phi) =
m^2\phi^2/2 + V_0$, but in Section \ref{Other} we discuss some other
theories with negative potentials.

In Section \ref{singularity} we will discuss cosmological
evolution near the initial and final singularities, and in particular
the role of particle production and anisotropy near the singularity.

In Section \ref{Cycles} we will apply our methods to the investigation
of the cyclic scenario.  As we will see, 
the cyclic regime in this scenario does not appear automatically. One should fine-tune the potential $V(\phi)$ and learn how to work with the super-Planckian potentials $|V(\phi)| >1$. One should also introduce superheavy particles with specific properties, study their production  at the singularity, and make sure that they do not affect the present stage of the evolution of the universe.
This adds new ``epicycles'' to this scenario, making it even more  
speculative. We discuss several possible modifications of this
scenario and conclude that the best way to improve it is to add a
usual stage of inflation before the stage of large scale structure
formation. This modification resolves many problems of the original
version of the cyclic scenario.  In this modified form of the
cyclic scenario inflation is once again the source of density
perturbations as well as the resolution of the cosmological problems
such as homogeneity and flatness.

Section \ref{Conclusions} summarizes our main  conclusions concerning cosmology with negative potentials and cyclic universe.

\section{\label{archetypes}Four Basic Regimes in Scalar Field Cosmology}

\subsection{\label{toymodel}A toy model with $V(\phi)= {1 \over 2}m^2\phi^2+ V_0$}

We will study the behavior of a homogeneous scalar field in a
Friedmann universe with the metric
\begin{equation}\label{metric}
ds^2=-dt^2+a^2(t)ds_3^2 \ ,
\end{equation}
where $ds_3^2=\gamma_{ij}dx^idx^j$ is the metric of a 3d space with
constant curvature, $k=0,\pm1$.

In this paper we will use a system of units in which $M_p = 1$, where
$M_p = (8\pi G)^{-1/2} \sim 2\times 10^{18}$ GeV.  The Friedmann
equation for a scalar field with potential energy density $V(\phi)$ is
\begin{equation}\label{f}
H^2= \left({\dot a\over a}\right)^2 ={1 \over 3 }\rho - {k \over a^2}
= {1 \over 3 } \left({1 \over 2}\dot \phi^2+V(\phi)
+\rho_\alpha\right) - {k \over a^2} \ .
\end{equation}
Here $\rho$ is the total energy density and $\rho_\alpha$ is the
density of matter with equation of state $p_\alpha = \alpha
\rho_\alpha$. For non-relativistic matter $\alpha = 0$, while for
radiation $\alpha = 1/3$.

The evolution of $H$ is given by a combination of the Einstein
equations
\begin{equation}\label{dot}
\dot H=-{1 \over 2}(\rho + p) + {k \over a^2} = -{1 \over 2}(\dot \phi^2 +
\rho_\alpha(1+ \alpha)) + {k \over a^2} \ .
\end{equation}
Alternatively, one can use the equation
\begin{equation}\label{ddot}
{\ddot a\over a}= - {1\over 6} (\rho + 3 p )= {1 \over 3}(V(\phi)-\dot
\phi^2) - {1\over 6} \rho_\alpha(1 + 3 \alpha ) \ .
\end{equation}
The evolution of the scalar field $\phi$ follows from the Einstein
equations,
\begin{equation}\label{field}
\ddot \phi + 3 H \dot \phi + V_{,\phi}=0  \ .
\end{equation}

We shall study the basic properties of scalar field cosmology using as
an example the simplest harmonic oscillator potential
\begin{equation}\label{basicpot}
V(\phi)={1 \over 2}m^2\phi^2 + V_0 \ ,
\end{equation}
Surprisingly, we will find that
cosmology with the potential (\ref{basicpot}) with $V_0 <0$ shares some common features with the cosmology of the ``inverse'' harmonic oscillator potential
\begin{equation}\label{basicpot1}
V(\phi)=-{1 \over 2}m^2\phi^2 -V_0 \ .
\end{equation}
 In particular, the expansion of the universe in theories with $V_0
< 0$ always turns into cosmological contraction.

Constructing phase portraits is a powerful method for investigating
the dynamics of the scale factor/scalar field system
(\ref{dot})-(\ref{field}).  Before we look at the phase portraits for
various values of $V_0$ in this model, it will be useful to discuss
some of their features. For the remainder of this section we will
consider $k=0$, i.e. flat universes. While this case will be the main
focus of our discussion throughout the paper, we will in several cases
refer to the extension of our results to open or closed universes as
well.

There are four basic regimes that we may encounter: the universe can
be dominated by the potential energy density $V(\phi)$, by the kinetic
energy density $\dot\phi^2/2$, by the energy density of an oscillating
scalar field, in which case $V(\phi)\sim \dot\phi^2/2$, or by
matter/radiation $\rho_\alpha$.

\subsection{The inflationary regime: Energy density dominated by $V(\phi)$}

Inflation occurs when the energy density is dominated by $V(\phi)$. In
this case $\dot\phi^2/2, \rho_\alpha \ll V(\phi)$ and $|\ddot \phi|
\ll |3H\dot\phi|$. This corresponds to the vacuum-like equation of
state
\begin{equation}\label{eqstate1}
p =  -\rho\ .
\end{equation}
The equations for $a$ and $\phi$ in this regime have the following
form:
\begin{equation}\label{dot11}
H^2 =\left({\dot a\over a}\right)^2 ={{m^2\phi^2 }\over 6 } +{V_0
\over 3 } \ ,
\end{equation}
\begin{equation}\label{field11}
 3 { \dot a \over a} \dot \phi + m^2\phi =0  \ .
\end{equation}
The solutions of the equations for $\phi(t)$ and $a(t)$ for the most
interesting case ${m^2\phi^2\over 2}\gg |V_0|$ are given by \cite{Chaot,book}
\begin{equation}
\label{1.7.23} \phi(t) =\phi_0  -  \sqrt{2\,\over 3} {m  t}  ~,
\end{equation}
\begin{equation}
\label{1.7.25} a(t)=a_0\,\exp\left(\frac{\phi_0^2-\phi^2(t)}{4}\right)\, ~.
\end{equation}
These solutions, which describe inflationary expansion, are valid only
for $\dot\phi^2/2 \ll V(\phi)$, which implies that inflation ends at
\begin{equation}\label{endin}
 |\phi_e| \sim 1  ~.
\end{equation}
In this paper we will assume that $m^2 \gg |V_0|$, in which case
${m^2\phi^2\over 2}\gg |V_0|$ is always satisfied during inflation.

Note that the same solution is valid if one reverses the time arrow,
$t \to -t$, in which case it describes a quasi-exponential contraction
of the universe (deflation).

\subsection{The kinetic regime: Energy density dominated by $\dot\phi^2/2$}

Another important regime occurs when the energy density is dominated
by $\dot\phi^2/2$. In this case $V(\phi), \rho_\alpha \ll
\dot\phi^2/2$ and $|\ddot \phi|, |3H\dot\phi| \gg m^2\phi$.  This
corresponds to the ``stiff'' equation of state
\begin{equation}\label{eqstate1a}
p =   \rho\ .
\end{equation}
The equations for $a$ and $\phi$ are:
\begin{equation}\label{dot11a}
H^2 =\left({\dot a\over a}\right)^2 ={{\dot\phi^2 } \over 6 } \ ,
\end{equation}
\begin{equation}\label{field11a}
 {\ddot\phi \over \dot\phi} = - 3 { \dot a \over a} \ .
\end{equation}

The solutions can be written as follows:
\begin{equation}\label{akin}
a(t) = t^{1/3} ~,
\end{equation}
\begin{equation}
\label{1.7.25a} \phi  = \phi_0 \pm \sqrt{2\over 3}\ \ln {t_0\over t} ;
\qquad {\dot\phi^2\over 2} = {1\over 3 t^2} \ .
\end{equation}
These solutions can describe an expanding universe or a universe
collapsing towards a singularity.

During the expansion of the universe, the inflationary regime $V(\phi)
\gg \dot\phi^2/2$ represents a stable intermediate asymptotic
  attractor. Even if a flat universe
begins in a state with $\dot\phi^2/2 \gg V(\phi) $, it typically
rapidly switches to an inflationary regime with $V(\phi) \gg
\dot\phi^2/2$ \cite{Khalat,KLS,Linde:ub}. This occurs because during
the expansion of the universe with $\dot\phi^2/2 \gg V(\phi) $, the
value of the kinetic energy drops down like $t^{-2}$, whereas the
field changes only logarithmically. Therefore for all power-law
potentials, the value of $V(\phi)$ decreases much more slowly than
$\dot\phi^2/2$. When it becomes greater than $\dot\phi^2/2$, inflation
begins.

During the collapse of the universe, the opposite occurs. $V(\phi)$
grows only logarithmically, whereas $\dot\phi^2/2$ diverges as $
t^{-2} $, where $t$ is the time remaining before the big crunch
singularity.  This means that the regime $\dot\phi^2/2 \gg V(\phi) $
generically occurs at the stage of collapse. In this regime one can
neglect $V(\phi)$ in the investigation of the singularity at $t \to
0$.

\subsection{The oscillatory regime: Evolution determined by the energy
density of an oscillating scalar field}

Now let us assume that the field $\phi$ oscillates near $\phi =0$ with
frequency much greater than $H$,  and that the average value of $V(\phi)$ during these oscillations is much greater than $V_0 = V(0)$. In this case one can neglect the term
$3H\dot \phi$ in Eq. (\ref{field}), so that in the first approximation
one simply has
\begin{equation}\label{field3}
\ddot \phi   + m^2\phi =0
\end{equation}
and
\begin{equation}\label{field3a}
 \phi  = \Phi~\sin mt   \ .
\end{equation}
Here $\Phi$ is the amplitude of the oscillation. The pressure $p =
\dot\phi^2/2 - V(\phi)$ produced by these oscillations is given by $
{m^2\over 2} \Phi^2~ \cos 2 mt$, so if one takes an average over many
oscillations, the pressure vanishes, $p \approx 0$. The universe in
this regime expands as $a \sim t^{2/3}$. Since the total energy of
pressureless matter is conserved, the amplitude of the oscillations
decreases, $\Phi(t) \sim a^{-3/2} \sim t^{-1}$.

The regime of oscillations usually begins after the end of inflation,
at $\phi \lesssim 1$. As long as one can neglect $V_0$, the field
oscillations after inflation approach the following asymptotic
regime \cite{KLSpreh}:
\begin{equation}\label{decr}
\phi(t) \approx {2\sqrt 2 \over \sqrt 3~mt} \, \sin \, mt ~\approx ~{
\sqrt 2 \over \pi \sqrt 3~N} \, \sin \, mt\ .
\end{equation}
Here $t$ is the time after the end of inflation and $N$ is the number of oscillations.

\

\FIGURE[!h]{\epsfig{file=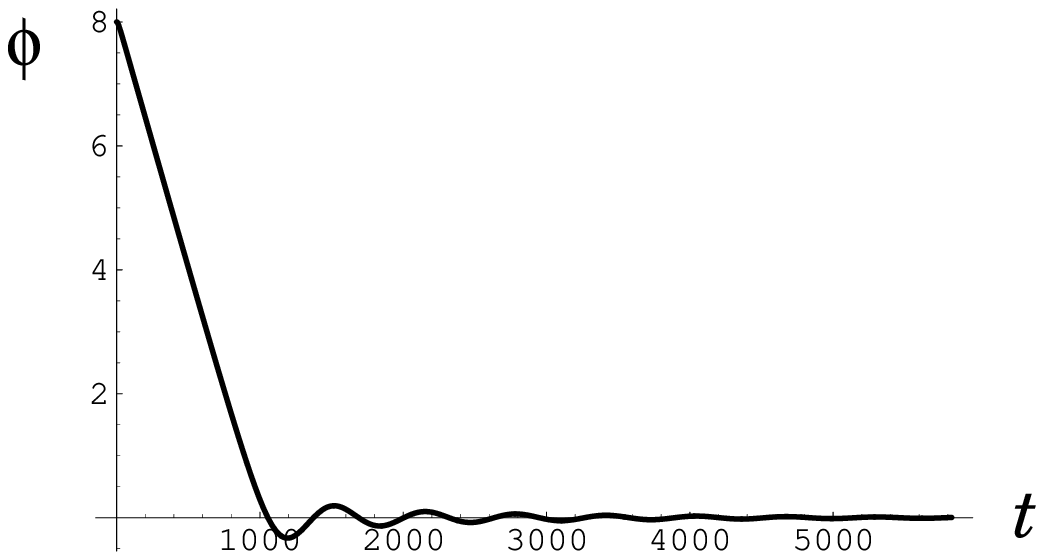,width=.47\columnwidth} \qquad
\epsfig{file=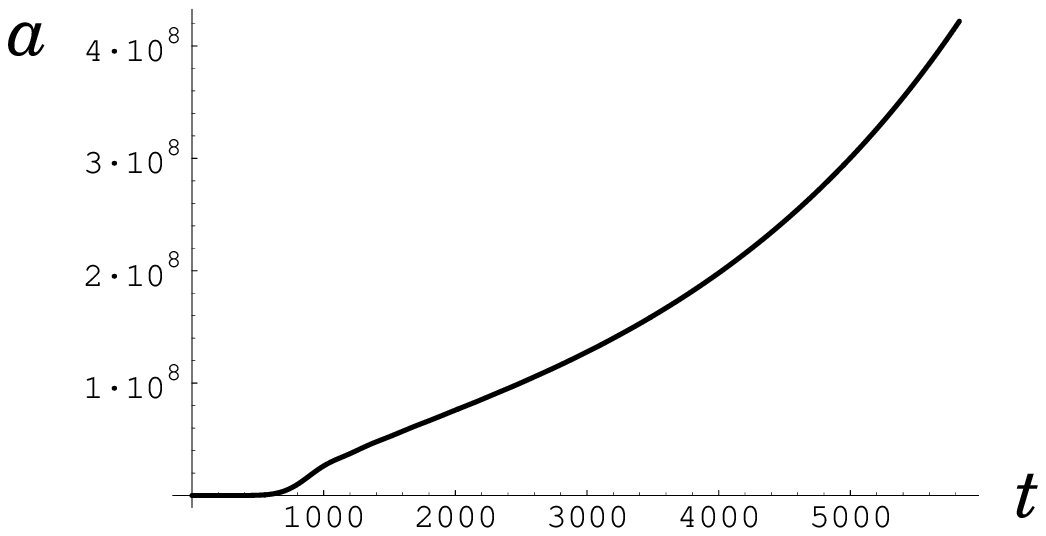,width=.47\columnwidth} \caption{Evolution of
the scalar field and the scale factor in the model $V(\phi) =
{m^2\over 2} \phi^2 + V_0 $ with $V_0 > 0$. In the beginning we have a
stage of inflation with the field $\phi$ linearly decreasing at $\phi
> 1$. At this stage the equation of state is $p \approx -\rho$. Then
the field enters a stage of oscillations with a gradually decreasing
amplitude of the field; $p \ll \rho$. When the energy of the
oscillations becomes smaller than $V_0$, the universe enters a second
stage of inflation.} \label{qfigplus}}

It is amazing that this simple model with $V_0>0$ can describe not only inflation in the early universe, but also the present stage of inflation/acceleration.
Indeed, when the amplitude becomes very small the term $V_0$ will become
important, and the universe enters a second stage of inflation
with $H^2 = V_0/3$.  The amplitude of oscillations of the field $\phi$
in this regime falls down exponentially. In particular, for $m^2\gg
H^2$ the amplitude decreases as $e^{-3Ht/2}$. The evolution of the
scalar field and the scale factor in the theory with $V_0 > 0$ is
shown in Fig. \ref{qfigplus}.

Meanwhile, if one considers the model with $V_0 <0$, a dramatic change occurs when the energy density of oscillations (and matter) gradually decreases and becomes
comparable to $-V_0$. According to Eqs. (\ref{f}) and (\ref{dot}), the
expansion of the universe slows down at that time, and eventually the
universe begins collapsing, see   Fig. \ref{qfig2}.

\FIGURE[!h]{\epsfig{file=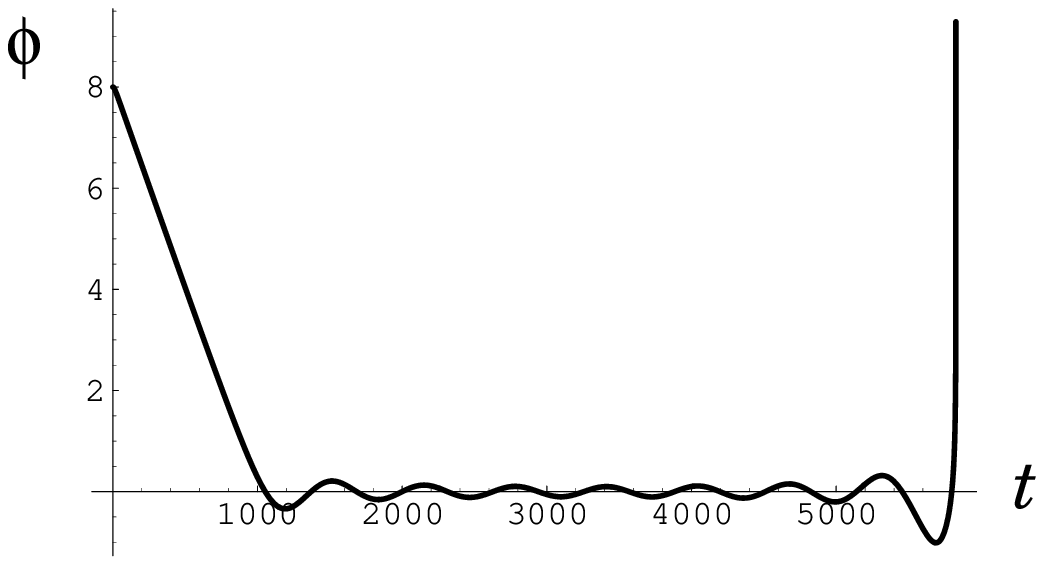,width=.47\columnwidth} \qquad
\epsfig{file=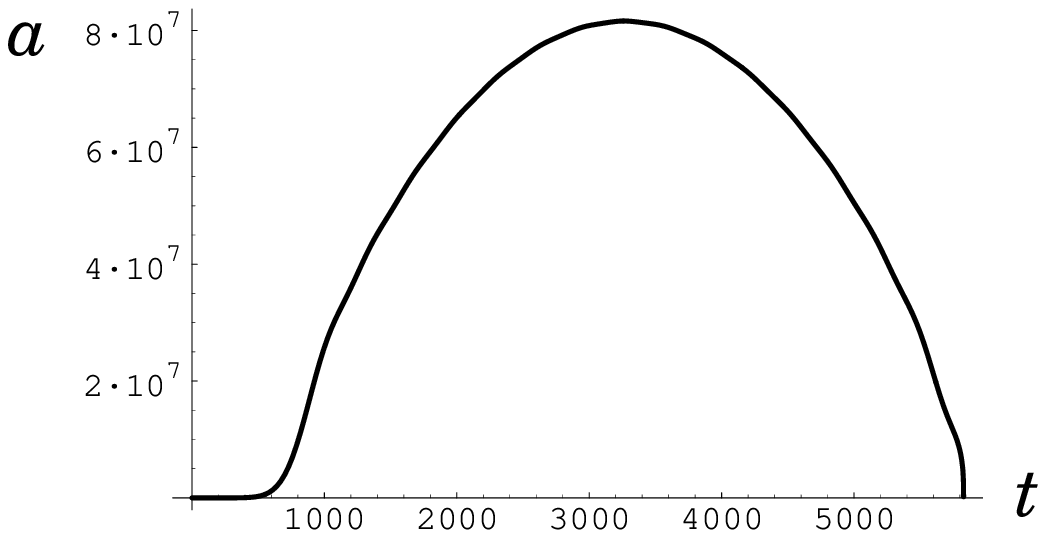,width=.47\columnwidth} \caption{Evolution of the
scalar field and the scale factor in the model $V(\phi) = {m^2\over 2}
\phi^2 + V_0 $ with $V_0 < 0$. In the beginning we have a stage of
inflation with the field $\phi$ linearly decreasing at $\phi > 1$. At
this stage the equation of state is $p \approx -\rho$. Then the field
enters a stage of oscillations with a gradually decreasing amplitude
of the field; $p \ll \rho$. When the energy of the oscillations
becomes equal to $\vert V_0\vert$, the universe stops expanding and
begins to contract. At this stage the amplitude of oscillations
grows. When it becomes greater than $O(1)$, the field stops
oscillating, the energy density is dominated by the kinetic energy of
the scalar field, $p \approx \rho$, and the universe collapses.}
\label{qfig2}}

When the universe contracts, the amplitude of oscillations grows as
$a^{-3/2}$. However, this process does not continue too long. Indeed,
let us compare $3H\dot\phi$ and $m^2\phi$ in this regime. If one can
neglect $V_0$ (and this is always the case for a sufficiently large
$\Phi$), one has $H \approx m\Phi/\sqrt 6$ and $\dot\phi \sim m
\phi$. Therefore one has $|3H\dot\phi| \gg |m^2\phi|$ for $\phi \gg
1$, so instead of Eq. (\ref{field3a}) one should use
Eq. (\ref{field11a}). Thus, during the collapse of the universe the
stage of oscillations ends and the regime dominated by kinetic energy
begins at
\begin{equation}\label{endosc}
|\phi_b| \sim 1 ~.
\end{equation}
Note that $|\phi_b| \sim |\phi_e|$, see Eq. (\ref{endin}). 

We will study the switch from expansion to contraction in a flat
universe in a much more detailed way in Section \ref{switch}. However,
we would like to make here some comments concerning this process.

The general textbook wisdom is that open and flat universes expand
forever, whereas closed universes eventually collapse. This lore was
based on investigation of universes with vanishing cosmological
constants. A closed universe with a sufficiently large positive
cosmological constant may expand forever, whereas open and flat
universes with a negative cosmological constant eventually collapse.

One of the well-known solutions of this type is an open universe with
a negative vacuum energy $V_0$. There is a solution to the Friedmann
equation $H^2 -a^{-2} = V_0/3$ for $V_0 <0$: ~ $a(t) = \sqrt{3\over
|V_0|}~\sin \sqrt{|V_0|\over 3}\, t $. This is a specific section of
anti de Sitter space, which is popular in M-theory and brane
cosmology. This universe has a coordinate singularity at $t = \pi
\sqrt{3\over |V_0|}$. Naively, one might think that this is exactly
what we have found in our investigation of universes with $V_0 <0$,
namely that when the energy density of matter in an expanding universe
decreases and the total energy density becomes dominated by a negative
cosmological constant, our universe reaches an AdS regime dominated by
a negative cosmological constant.

However, this is not the case. We discuss here a flat universe regime,
which appears after a long stage of inflation. In this case (unless
one considers open inflation models with $\Omega <1$) the term
${k\over a^{2}}$ with $k = \pm 1,0$ can be omitted in the general
Friedmann equation. The Friedmann equation $H^2 = \rho/3$ describing a
flat universe does not have any solutions with $\rho(\phi)< 0$. Once
the universe approaches the turning point where the total energy
density vanishes it begins collapsing, and the total energy density
becomes positive again \cite{Banks:1995dt,N8,Fastroll}. Thus the
standard inflationary prediction $\Omega = 1$ implies that we cannot
live in AdS space dominated by a negative cosmological constant
\cite{N8,Fastroll}.

\subsection{\label{radiation}Evolution determined by the energy density
of matter or radiation}

The first models of inflation were based on the assumption that the
universe from the very beginning was in a state of thermal
equilibrium; inflation began when the temperature of the universe
became much smaller than the Planck temperature $T \sim M_p$
\cite{Old,New}. Later it was found that this assumption is not
necessary, and in many models inflation may start immediately after
the big bang \cite{Chaot}. In this case the existence of matter prior
to inflation becomes less important, and sometimes it even hampers the
development of inflation \cite{book}. Therefore many works on initial
conditions for inflation neglect the possible impact of matter on the
motion of the scalar field and concentrate on finding self-consistent
cosmological solutions describing scalar fields in otherwise empty
universes. This is the simplest approach, especially in cases where
$\dot\phi^2/2 \ll V(\phi)$ and inflation begins immediately after the
big bang.

However, in some cases the scalar field initially may have large
kinetic energy, $\dot\phi^2/2 \gg V(\phi)$. Moreover, one may expect creation of relativistic  or non-relativistic
particle near the singularity. Note that the
existence of even a small amount of matter may have an important
effect on the motion of the field. Indeed, the kinetic energy of the
scalar field $\dot\phi^2/2$ in the regime $\dot\phi^2/2 \gg V(\phi)$
decreases as $a^{-6}$. Meanwhile, the density of radiation decreases
as $a^{-4}$ and the density of non-relativistic matter decreases as
$a^{-3}$. Therefore the energy density of matter eventually becomes
greater than $\dot\phi^2/2$. As we will see, once it occurs, the field rapidly slows down or even completely
freezes. This effect may provide good initial conditions for a
subsequent stage of inflation \cite{Toporensky:1999pk}.

Indeed, let us assume that in the beginning the field $\phi$ moves
very fast, so that $|3H\dot\phi| \gg |V,_\phi| = |m^2\phi|$. Suppose,
however, that at some moment the energy density of the universe
becomes dominated by matter with the equation of state $p_\alpha =
\alpha \rho_\alpha$. In this regime one can represent the cosmological
evolution in the following form \cite{book}
\begin{eqnarray}\label{alpha}
\rho_\alpha &=& \rho_\alpha(t_0) \left({a(t)\over
 a_0}\right)^{-3(1+\alpha)}, \nonumber \\ a(t) &=& a_0\left({t\over
 t_0}\right) ^{2\over 3(1+\alpha)}, \nonumber \\ H &=& {2\over
 3(1+\alpha) t}~, \nonumber \\ \dot\phi &=& \dot\phi_0 {a_0^3\over
 a^3} = \dot\phi_0 \left({t_0\over t}\right)^{2\over 1+\alpha}~.
\end{eqnarray}
This regime has a very interesting feature: Even if it continues for
an indefinitely long time, the change of the field $\phi$ during this
time remains quite limited.  Indeed,
\begin{equation} \label{changephi}
\Delta\phi \leq \int\limits_{t_0}^\infty \dot\phi dt = \dot\phi_0
\int\limits_{t_0}^\infty \left({t_0\over t}\right)^{2\over 1+\alpha}
dt = {1+\alpha\over 1-\alpha}~\dot\phi_0 t_0 .
\end{equation}
If $t_0$ is the very beginning of matter domination ($\dot\phi_0^2/2
\sim \rho_\alpha$), then $\dot\phi_0 t_0 \sim {2\over \sqrt 3(
1+\alpha)} = O(1)$. Therefore
\begin{equation}\label{rchange}
\Delta\phi \lesssim 1  
\end{equation}
in Planck units (i.e. $\Delta\phi \lesssim M_p$).  This means, in
particular, that a free field $\phi$ in a matter dominated universe
cannot move by more than $O(M_p)$.

This simple result has important implications.  In particular, if the
motion of the field in a matter-dominated universe begins at $|\phi|
\gg 1$, then it can move only by $\Delta\phi \lesssim 1$. Therefore in
theories with flat potentials the field always remains frozen at
$|\phi| \gg 1$.

The field begins moving again only when the Hubble constant decreases
and $|3H\dot\phi|$ becomes comparable to $|V,_\phi|$. But in this case
the condition $ 3H\dot\phi \approx |V,_\phi|$ automatically leads to
inflation in the theory $m^2\phi^2/2 +V_0$ for $|V_0| < m^2$ and $\phi
\gg 1$.

This means that even a small amount of matter or radiation may
increase the chances of reaching a stage of inflation, see   \cite{Toporensky:1999pk} and Fig. \ref{comparison} in Section \ref{negative}. Indeed, consider any theory with $V(\phi) \sim \phi^n$. Suppose in the
beginning we had a kinetic energy dominated regime $\dot\phi^2/2 \gg
\rho_\alpha, V(\phi)$ starting at $\phi \gg 1$. Then the field $\phi$
would change very slowly, whereas $\dot\phi^2/2$ would rapidly drop
down until it became comparable either to $V(\phi)$ or to
$\rho_\alpha$. If at that moment $V(\phi) > \rho_\alpha$, inflation
would begin immediately. But even in the most unfavorable case
$V(\phi) \ll\rho_\alpha$ inflation would begin eventually. Indeed, at
$\phi \gg 1$ one has the double inequality $m^2 = V'' \ll V(\phi) \ll
\rho_\alpha \sim H^2$. Therefore the Hubble constant is much greater
than the effective scalar field mass. In this case the field
practically does not move until the desirable regime $V(\phi) >
\rho_\alpha$ is reached and inflation begins.

\section{\label{portraits}Phase Portraits and Cosmological Evolution}

Having discussed some important limiting regimes in scalar field
cosmology, we are now ready to investigate the complete evolution of a
Friedmann universe with a scalar field. Later we will discuss the
effects of adding matter to this system, but for now we restrict
ourselves to a system with three independent variables, $\phi$,
$\dot{\phi}$, and $H$. To study this system we find it most convenient
to rewrite the evolution equations for $a$ and $\phi$ as a set of
three coupled, first-order, differential equations:
\begin{eqnarray}
\label{xdot}{d \phi \over dt} &=& \dot{\phi} \\
\label{ydot}{d \dot{\phi} \over dt} &=& -3 H \dot{\phi} - V_{,\phi} \\
\label{zdot}{d H \over dt} &=& -{1 \over 3} \left(\dot{\phi}^2 - V\right) - H^2
\end{eqnarray}
plus the constraint equation
\begin{equation}
\label{xyzconstraint}H^2 - {1 \over 6} \dot{\phi}^2 - {1 \over 3}V =
-{k \over a^2}.
\end{equation}

All solutions to these three equations can be represented as
trajectories in the 3d phase space of $\phi$, $\dot{\phi}$, and
$H$. Simply looking at plots showing a number of these trajectories
can help give some intuition for the cosmology of a particular model
(as defined by the potential $V$). There are a number of ways to get
more information out of the phase portraits, however.

One important step is to determine all of the critical points,
i.e. the points for which the derivatives of all three phase variables
vanish. There are finite and infinite critical points.
 Every trajectory must begin and end  at these
critical points.

To find infinite critical points and visualize the flow of trajectories at infinity,
a  useful trick is to do a Poincar\'{e} mapping
\begin{equation}
x_P \equiv {x \over 1+r}\ ,
\end{equation}
where $x$ is any of ($\phi$, $\dot{\phi}$, $H$) and $r^2 = \phi^2 +
\dot{\phi}^2 + H^2$. The interior of the unit sphere $\phi_P^2 +
\dot{\phi}_P^2 + H_P^2 = 1$ maps to the infinite phase space of
$\phi$, $\dot{\phi}$, and $H$, so by plotting trajectories in these
new coordinates the entire phase space can be easily visualized. At
times in this paper we will plot a 2d phase portrait, e.g. in the
variables $\phi$ and $\dot{\phi}$ only. In these cases we use a 2d
Poincar\'{e} mapping where $r^2 = \phi^2 + \dot{\phi}^2$.

With the Poincar\'{e} mapping it is possible to identify a set of
infinite critical points, namely those that occur on the bounding
sphere $\phi_P^2 + \dot{\phi}_P^2 + H_P^2 = 1$. These points represent
the possible starting and ending points for all trajectories that go
off to infinity in the usual coordinates.

Because no two trajectories can ever cross in phase space, it is easy
to define the behavior of a system whose phase portrait is two
dimensional. Fortunately, for the cosmological systems we are
considering we can identify a 2d surface that separates different
regions of the 3d phase space.  For the flat universe $k=0$ the
constraint equation (\ref{xyzconstraint}) defines a 2d surface.  All
trajectories in this case are located at this surface, i.e. the phase
portrait for the flat universe is two dimensional.  This surface in
turn divides the phase space into three separate regions (including
the surface itself) representing the possible types of curvature. No
trajectory can pass from one of these regions to another.
Although the location of the finite critical points for a
given model depends strongly on $V$, the structure of the infinite
critical points is very similar across a wide range of potentials. See
\cite{Felder:2001da} for recent discussion.

\section{\label{positive}Cosmology with a Non-Negative Potential}

As a simple example we consider the model $V(\phi)=V_0+{1 \over
2}m^2\phi^2$ discussed in Section \ref{toymodel}. By rescaling the
field and time variables the mass $m$ can be eliminated from the
equations, so for simplicity we simply set $m=1$ in what follows. Thus
the evolution and constraint equations become
\begin{eqnarray}
{d \phi \over dt} &=& \dot{\phi} \\
{d \dot{\phi} \over dt} &=& -3 \dot{\phi} H - \phi \\
{d H \over dt} &=& -{1 \over 3} \dot{\phi}^2 + {1 \over 6} \phi^2 + {1 \over 3} V_0 - H^2
\end{eqnarray}
\begin{equation}
6 H^2 - \dot{\phi}^2 - \phi^2 - 2 V_0 = -6 {k \over a^2}.
\end{equation}

The hypersurface representing a flat universe is given by setting
$k=0$ in the constraint equation, which gives
\begin{equation}
6 H^2 - \dot{\phi}^2 - \phi^2 = 2 V_0.
\end{equation}
The surface defined by this equation is a hyperboloid. For positive
definite potentials $V_0>0$ it is a hyperboloid of two sheets, meaning
the two branches at $H>0$ and $H<0$ are disconnected. For $V_0=0$ this
hyperboloid reduces to a double cone.

There are two finite critical points for this system at $\phi=\dot{\phi}=0$,
$H=\pm \sqrt{V_0/3}$. For $V_0=0$ these two points reduce to a single
finite critical point at the origin. To find the infinite critical
points we first rewrite the evolution equations in terms of the
Poincar\'{e} variables and then set their derivatives equal to
zero. This yields eight points.

Figure \ref{phase1} shows the phase space for this model with $V_0>0$
along with a sample of trajectories for $k=0$. The hyperboloid along
which all of these trajectories lie represents a flat universe.  
The upper branch corresponds to expansion and the lower one to
contraction. The fact that the two branches are disconnected means
that in a flat universe in this model expansion can never reverse and
become contraction. Note that this conclusion is unchanged for the
case $V_0=0$. In that case the hyperboloid becomes a double cone and
the two branches touch at a single point. Since that point is a
critical point, however, no trajectories can pass from one branch of
the cone to the other. The lower branch corresponds to the upper
branch with time reversal $t \to -t$. The upper branch of the flat
universe hyperboloid is shown projected into a 2d plot in Figure
\ref{phase1a}.\footnote{This is not a direct ``shadow'' since it uses
the 2d rather than the 3d Poincar\'{e} mapping, see Section
\ref{portraits}. Effectively the upper branch of the hyperboloid is
stretched out onto the circle rather than vertically projected down to
it. From here on we will refer to such 2d portraits as projections of
the 3d ones.} This plot is very similar to the one shown in
\cite{Khalat} for this model with $V_0=0$.

\FIGURE[!h]{\epsfig{file=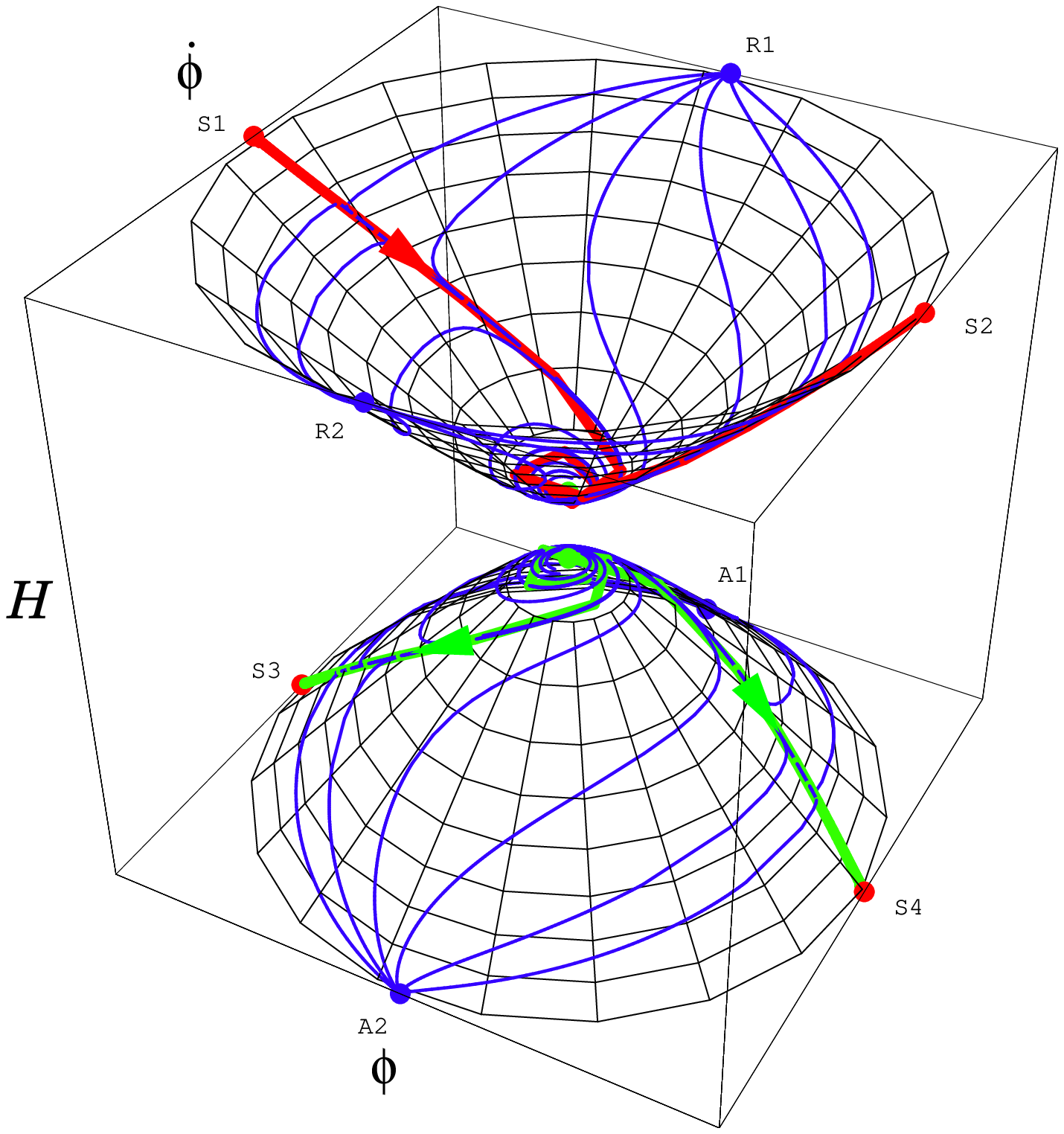,width=.9\columnwidth}
\caption{Phase portrait for the theory
$V(\phi)={1 \over 2}m^2\phi^2+ V_0$ with $V_0>0$ in rescaled
coordinates $(\phi,\dot\phi,H)$. The branches describing stages of
expansion and contraction (upper and lower parts of the hyperboloid)
are disconnected.} \label{phase1}}

\FIGURE[!h]{\epsfig{file=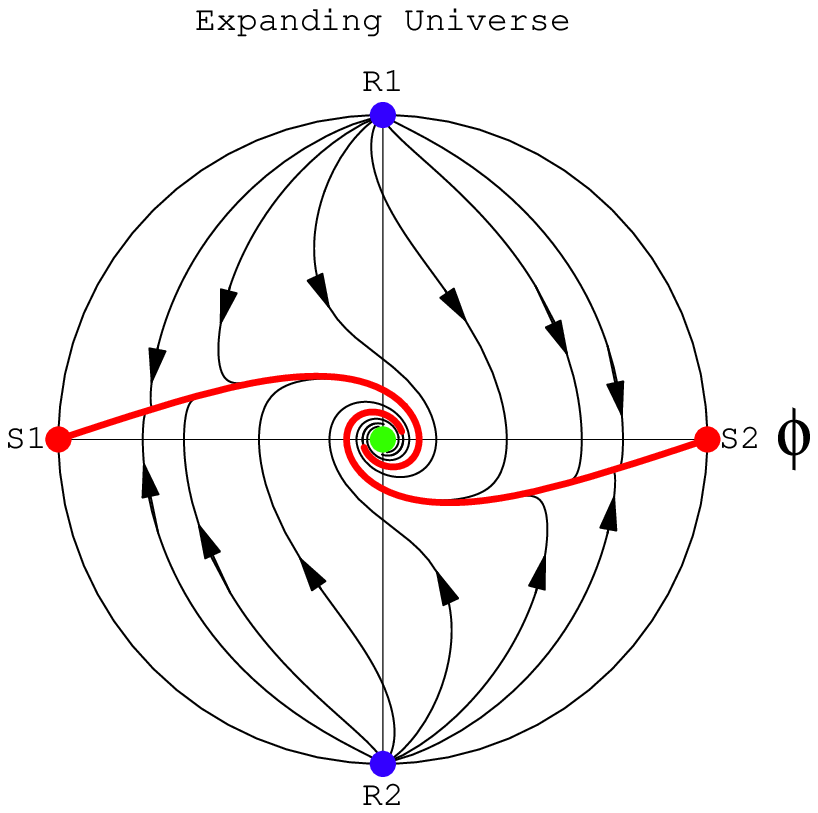,width=.6\columnwidth}
\caption{Projection of the upper branch of the full phase portrait for
the theory $V(\phi)={1 \over 2}m^2\phi^2+ V_0$ with $V_0>0$ in
rescaled coordinates $(\phi,\dot\phi)$.} \label{phase1a}}

For an expanding universe there are four infinite critical points, two
repulsors labeled $R_1$ and $R_2$ and two saddle points labeled $S_1$
and $S_2$. All trajectories begin at $R_1$,  $R_2$  and wind
towards the focus at the center.  The separatrices emanating from $S_1$
and $S_2$ represent attractor trajectories (not to be confused with
attractor critical points). Along these trajectories the universe
experiences inflation ($\phi^2 \gg \dot{\phi}^2$) until it nears the
center and begins winding around it, corresponding to field
oscillations near the potential minimum. These separatrices represent
a set of measure zero in the space of trajectories; the two shown are
the only trajectories that begin at the saddle points. Nonetheless
they are important because most of the trajectories emanating from the
repulsor points asymptotically approach the separatrices. This is why
inflation is a generic feature of models such as this one, and also
why inflation erases all information about the initial conditions that
preceded it.

Thus a typical trajectory passes through three of the four regimes
described in Section \ref{archetypes}. Near the repulsors the kinetic
energy dominates and the equation of state is stiff, $p \approx
\rho$. Near the main part of the separatrices the equation of state is
inflationary, $p \approx -\rho$. Finally near the center the scalar
field oscillates and the equation of state is that of non-relativistic
matter, $p \ll \rho$. During the oscillations  the scalar field
decreases as
\begin{equation}\label{decr2}
\phi(t) \approx {1 \over {4 N}} \, \sin \, mt \ , 
\end{equation}
where $N$ is the number of oscillations, see Eq. (\ref{decr}). Although particle production
is not included in these phase portraits, this evolution will
typically end with the scalar field decaying into other forms of
matter, thus finishing the evolution in the fourth regime, matter
and/or radiation domination. The contracting branch is a mirror image
of the expanding one, with the same three regimes occurring in the
opposite order, finally ending with a big crunch singularity at the
attractor points $A_1$ and $A_2$.

For an open or closed universe the trajectories would lie in the
interior or exterior of the hyperboloid, respectively  \cite{Khalat}. For an open
universe nearly all trajectories would asymptotically approach the
separatrices on the flat universe hypersurface. This tendency reflects the
fact that for most initial conditions inflation will occur and drive
the universe towards flatness. Once this has occurred the trajectories
spiral in towards the focus at the bottom of the hyperboloid. For a
closed universe there are also many trajectories that rapidly approach
these separatrices, but there is also a class of trajectories that
moves from the repulsive critical points to the attractive ones
without ever passing near the flat universe hypersurface. These
trajectories reflect closed universes that collapse rapidly before
inflation has a chance to occur.

\vskip -1cm
 
\hskip 1cm\FIGURE[!h] {\centerline{\epsfig{file=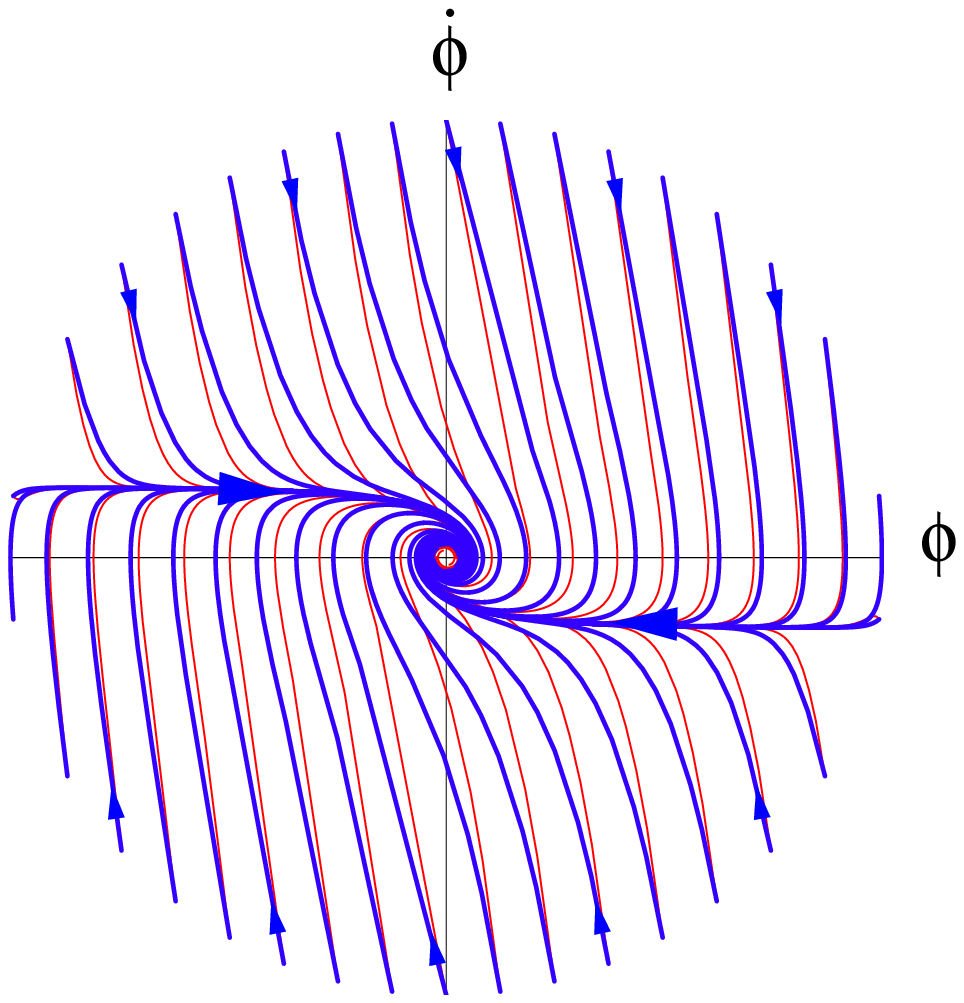,width=.6\columnwidth}}

\\

\caption{Phase portrait for the theory $V(\phi)={1 \over 2}m^2\phi^2$ without Poincar\'{e} mapping.  The blue (thick) lines show trajectories describing the universe without radiation. The scalar field has half Planck density at the beginning of the simulations. The red (thin) lines show trajectories where an
equal amount of energy in radiation was added to the system. As we
see, in the presence of radiation the velocity of the scalar field
rapidly decreases, which usually leads to the onset of inflation. }
\label{comparison}}

This conclusion becomes even more apparent if one takes into account
matter/radiation \cite{Toporensky:1999pk}. As we have argued in Section \ref{radiation}, the
existence of matter rapidly freezes the motion of the scalar
field. Therefore if the field $\phi$ was initially large and had a
large velocity such that $\phi \gg 1$, $\dot\phi^2/2 \gg V(\phi)$,
then the presence of matter would increase the probability of
inflation. This can be confirmed by comparing the phase portraits of
the universe with and without radiation. Although the phase portrait with radiation is three dimensional,
it is convenient to make its projection to the $(\dot \phi, \phi)$ plane,
 see Fig. \ref{comparison}.

In the second and fourth quadrants of this figure the field starts out
moving towards the minimum. The presence of radiation slows the field
down, causing it to move more quickly towards the inflationary
separatrix trajectory. In the first and third quadrants where the
field starts out moving away from the minimum the duration of
inflation is slightly diminished by the presence of radiation, but the
probability of inflation is nearly unity.

\section{\label{negative}Cosmology with a Negative Potential}

Now we turn to the main subject of our investigation, cosmological models with scalar field potentials that may become negative. We will continue using the simple example $V(\phi)=V_0+{1 \over 2}m^2\phi^2$, but now we will consider $V_0<0$. The hypersurface representing a flat universe is still defined by
\begin{equation}
6 H^2 - \dot{\phi}^2 - \phi^2 = 2 V_0,
\end{equation}
but with $V_0$ negative, this surface is a hyperboloid of one sheet.

Figure \ref{phase2} shows the phase space for this model and sample
trajectories for a flat universe. The phase space is two dimensional,
but its topology is very different from that for non-negative
potentials. The infinite critical points are unchanged because the
finite term $V_0$ has no effect at infinity, but there are no finite
critical points. Thus all trajectories begin at infinity with $H>0$
and end at infinity with $H<0$. This is possible because the regions
corresponding to expansion and contraction are now connected.  This
property is valid for all types of curvature $k$, i.e. for open, flat
or closed universes.

\FIGURE[!h]{\epsfig{file=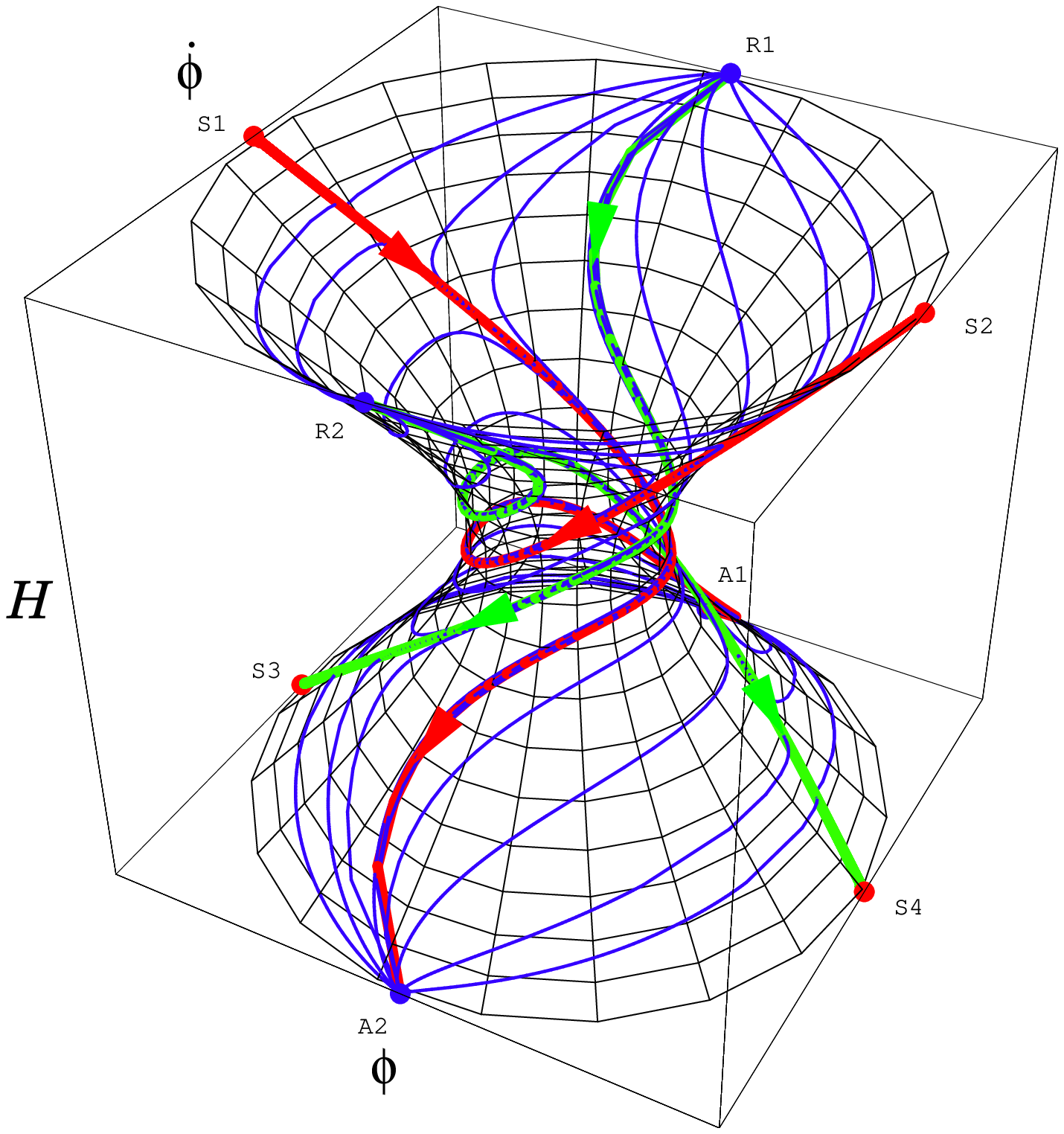,width=.9\columnwidth}
\caption{Phase portrait for the theory  $V(\phi)= {1 \over
2}m^2\phi^2+ V_0$ for $V_0<0$. The branches describing stages of
expansion and contraction (upper and lower parts of the hyperboloid)
are connected by a throat.} \label{phase2}}

To show a 2d projection of the flat universe hypersurface for this
model, we have to plot both the expanding and contracting branches, as
depicted on Figure \ref{phase3}. Trajectories in the expanding
universe region spiral in towards the center. When they touch the
inner circle, the ``throat'' of the hyperboloid, they pass into the
contracting universe region. There they spiral back out to infinity,
i.e. the big crunch. Thus typical trajectories in this scenario pass
through the three regimes described above, kinetic energy domination,
potential energy domination, and oscillations, and then pass back
through them in reverse order. As before, including particle
production will typically introduce a matter/radiation dominated
regime after the first stage of oscillations. Eventually, however, the
matter and radiation will redshift away and the universe will begin
contracting. We will examine this process in more detail in the next
section.

\EPSFIGURE[!h]{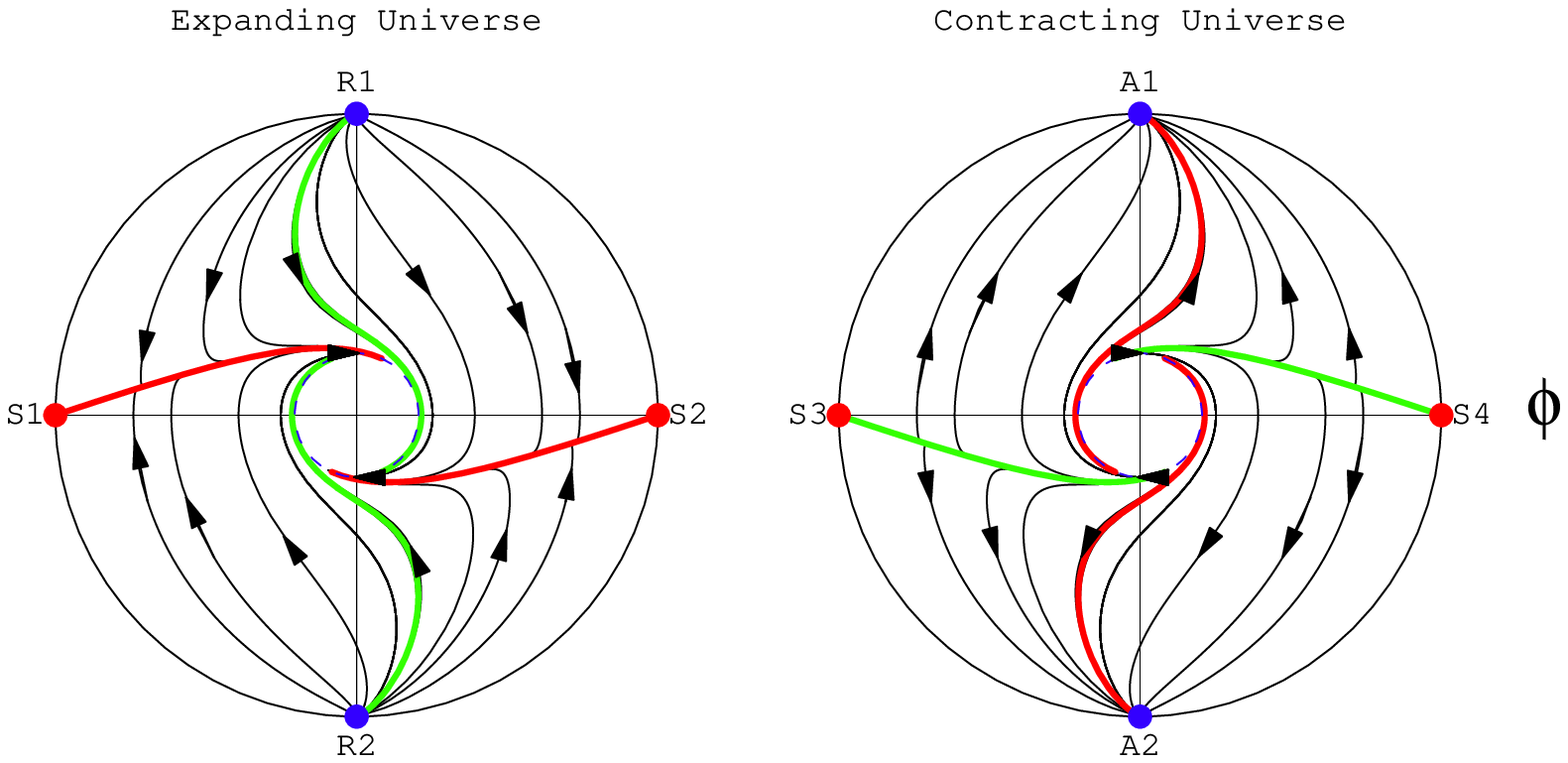}{Left: $(\phi,\dot\phi)$ projection of the
$H >0$ branch. Right: $(\phi,\dot\phi)$ projection of the $H <0$
branch. Trajectories from the left panel continue on the right
panel.\label{phase3}}

Aside from this ``wormhole'' connecting the expanding and contracting
branches this phase portrait looks a lot like the one for $V_0>0$
shown in Figure \ref{phase1a}. Note, however, that in this case the
separatrices emanating from the saddle points $S_1$ and $S_2$ no
longer spiral in to the center, but rather end up reaching the points
$A_1$ and $A_2$. Likewise there are separatrices
  that begin at $R_1$
and $R_2$ and end on $S_3$ and $S_4$. In the expanding phase their segments and segments of
nearby trajectories represent the rare cases that manage to avoid inflation. In the contracting phase they become the marginal
trajectories separating those that end at positive and negative
$\phi$. The number of windings (i.e. field oscillations) can be
estimated by setting $m^2 \phi^2/2=\vert V_0\vert$ and using
(\ref{decr2}) to give 
\begin{eqnarray}\label{number} N \approx {{m } \over
{6\sqrt{|V_0|}}} \ .  \end{eqnarray} 
(This number of windings can be used to
determine which repulsors and attractors are connected to which saddle
points, e.g. whether the separatrix that begins at $R_1$ ends at $S_3$
or $S_4$.)

\FIGURE[!h]{\epsfig{file=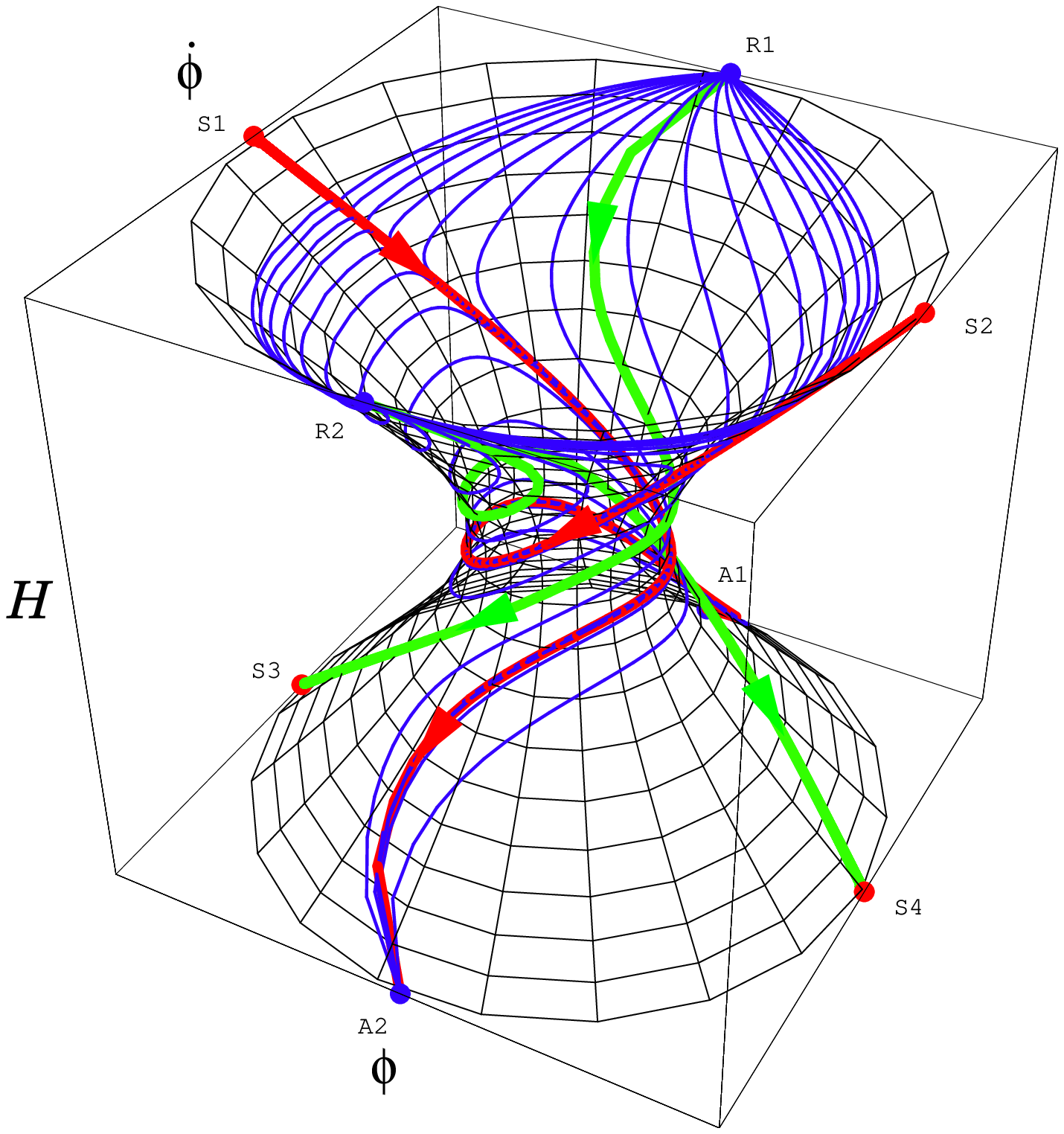,width=.9\columnwidth}
\caption{A different version of the phase portrait for the theory $V(\phi)= {1 \over 2}m^2\phi^2+ V_0$ for $V_0<0$. We begin with the trajectories evenly distributed with respect to the initial values of $\phi$ in the early universe (upper part of the hyperboloid) and see what happens to them in the lower part. These trajectories are concentrated near the red separatrices and repulsed from the green ones.} \label{phase2new}}

The phase portraits shown above were constructed in a way symmetric
with respect to time reversal, $t\to -t$. This is a legitimate
approach, since our equations allow all of the solutions shown in the
previous figures. However, one can obtain some additional information
if, for example, one considers trajectories equally distributed with
respect to the initial value of the field $\phi$ at the Planck time
and follows their evolution from the region with $H>0$ to the region
with $H<0$.

If we do so, the phase portrait shown in Fig. \ref{phase2} starts
looking somewhat different. Almost no trajectories beginning in the
upper part of the hyperboloid are seen in its lower part, and those
few that can be seen there are positioned very close to the (red)
separatrices going from $S1$ to $A2$, and from $S2$ to $A1$ see
Fig. \ref{phase2new}.  No trajectories are seen near the (green) lines
going from $R1$ to $S3$ and from $R2$ to $S4$.  This might seem
surprising because these lines are solutions of the equations of
motion, so there must be other solutions nearby. Indeed we have seen
them in Fig. \ref{phase3}.  However, the (red) lines going from
$S1$ to $A2$ and from $S2$ to $A1$ are strong attractors in the regime
$H<0$, whereas the lines going from $R1$ to $S3$ and from $R2$ to $S4$
are strong repulsors. Therefore most of the trajectories originating
at $H>0$ and homogeneously distributed with respect to the field
$\phi$ at the Planck density are repelled from the lines going from
$R1$ to $S3$ and from $R2$ to $S4$, and tend to merge with the lines
going from $S1$ to $A2$ and from $S2$ to $A1$.

This effect is especially apparent in the 2d phase portrait, where we
do not make the Poincar\'{e} mapping, see Fig. \ref{phase3new}. Most of
the trajectories coming from the panel with $H>0$ have merged with the
red separatrix on the panel corresponding to $H<0$.

\EPSFIGURE[!h]{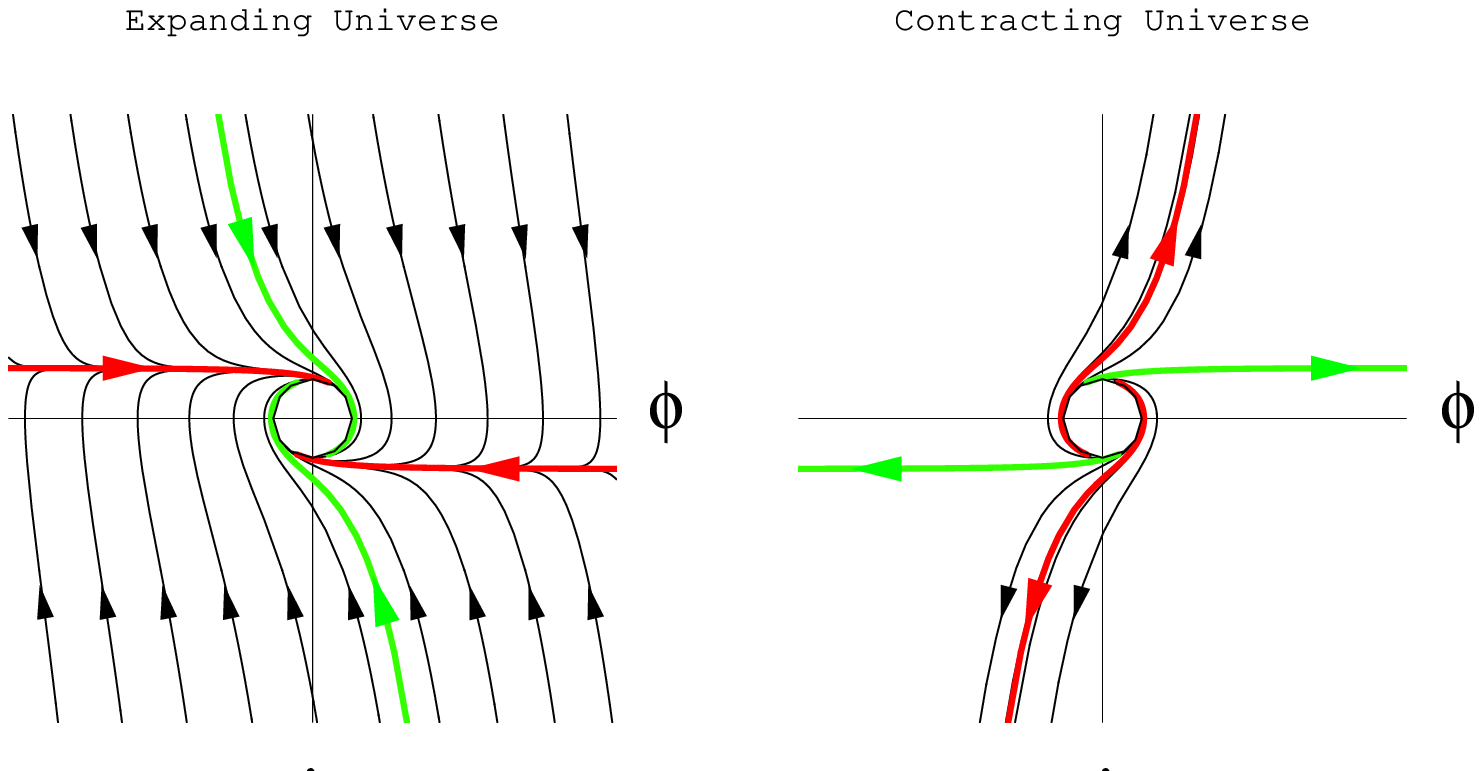}{As in the previous figure, we begin with the trajectories evenly distributed with respect to the initial values of $\phi$ in the early universe. However, now we show a 2d projection of these trajectories, without Poincar\'{e} mapping. Left: $(\phi,\dot\phi)$
projection of the $H >0$ branch. Right:
$(\phi,\dot\phi)$ projection of the $H <0$ branch. Trajectories from
the left panel continue on the right panel.\label{phase3new}}

An important (and obvious) feature of the 3d phase portraits
Fig. \ref{phase2} and  Fig. \ref{phase2new} is that the
separatrices, as well as other trajectories, never intersect in
3d. This is a trivial consequence of the fact that we are solving a
system of 3 first order equations for 3 variables, $\phi$, $\dot\phi$
and $H$. One of implications of this fact is that a bunch of
trajectories in the immediate vicinity of the (green) lines going from
$R1$ to $S3$ and from $R2$ to $S4$ never reach the inflationary regime
described by the (red) inflationary separatrices going from $S1$ to
$A2$ and from $S2$ to $A1$. Only the trajectories that are
sufficiently far away from the green lines going from $R1$ to $S3$ and from $R2$ to $S4$ can enter the stage of
inflation.

This observation will be important for us when we describe the cyclic
scenario \cite{Cyclic}, see Section \ref{Cycles}. In this regime the
red inflationary separatrices reach the singularity and are supposed to
bounce back. In the language of the phase portraits this bouncing back
implies that the end of the red line going from $S1$ to $A2$ becomes
the beginning of the green line going from $R1$ to $S3$. But in this
case the universe cannot attain the inflationary regime, since the
trajectories close to the green line never switch to the vicinity of
the red line.  Thus the cyclic regime is possible only if bouncing
from the singularity shifts the trajectory to the right from the green
separatrix. From Fig. \ref{phase3new} it is obvious that this shift
may happen either due to an increase of $\dot\phi$ or due to an
increase of the field $\phi$.

The evolution of this system in an open or closed universe is not very
different from the flat universe evolution, although the phase space
is three dimensional. Because of the structure of the trajectory flow
between their ends at the infinite critical points, all trajectories
pass from expansion to contraction, even for an open universe. As with
$V_0 > 0$ the trajectories for the open and closed cases will tend to
asymptotically approach the flat universe hypersurface, and more
specifically will tend to approach the inflationary separatrices. As
before, however, the closed universe will include some trajectories
that quickly collapse before experiencing inflation.

It is instructive to estimate the time that the universe may spend in
its post-inflationary expanding phase before it begins to contract.
The energy density of the oscillations of the scalar field, just like
the energy density of nonrelativistic matter, decreases as 
$\rho_{\rm CDM} \sim {4\over 3 t^2} $. The universe begins to collapse at
$\rho_{\rm CDM}+V_0 =0$. This happens at $t \sim {2\over\sqrt{
3|V_0|}}$. As one could expect, this time can be greater than the
present age of the universe only if $|V_0| \lesssim 10^{-120}$.

This estimate remains true for a wide variety of potentials and for
matter with any reasonable equation of state.  However, in the
theories where $V(\phi)$ has a very flat plateau or a local minimum,
the universe may spend a very long time before the field $\phi$ falls
down to the minimum with $V(\phi)<0$
\cite{N8,Fastroll,Cyclic}. Therefore in general the life-time of the
universe may be very large even in theories with a very deep minimum
of $V(\phi)$.

\section{\label{switch} Going from expansion to contraction in the
model $V(\phi) = {m^2\over 2}\phi^2 +V_0$}

Having analysed general properties of phase portraits in the theory
$V(\phi) = {m^2\over 2}\phi^2 +V_0$, let us study in a more detailed
way the most interesting feature of the models with $V_0 < 0$, the
switch from expansion to contraction. It is always possible to study
this process numerically, but sometimes one can do better than that.

It will be convenient to represent $V(\phi) = {m^2\over 2}\phi^2 +V_0$
in the form 
\begin{eqnarray}\label{simplepot} V(\phi) = {m^2\over 2}(\phi^2
-\phi_0^2) \ .  
\end{eqnarray} 
This potential has a minimum at $\phi = 0$, where
it takes a negative value $V(\phi) = -{m^2\over 2}\phi_0^2$. The
potential vanishes ($V(\phi) = 0$) at $\phi = \pm \phi_0$.

Let us assume, in the first approximation, that the scale factor of
the universe does not change much during each oscillation of the field
$\phi$. In such a case the field $\phi$ would experience a simple
oscillatory motion, 
\begin{eqnarray}\label{osc} 
\phi(t) = \Phi ~\cos mt \ , 
\end{eqnarray}
where $\Phi$ is the amplitude of the oscillations. In this case the
total energy density of the scalar field would remain constant, $\rho
= {m^2\over 2}(\Phi^2 -\phi_0^2)$.

This approximation works well for $\Phi \approx \phi_0$.
For $\Phi > \phi_0$, there are two cosmological solutions, describing
either an expanding universe with $H = +m \sqrt{(\Phi^2 -\phi_0^2)/6}$
or a contracting universe with $H = -m \sqrt{(\Phi^2 -\phi_0^2)/6}$.

If the Hubble constant $H$ is positive, the amplitude of the field and
its total energy density decrease. If the initial amplitude of the oscillations is much
greater than $\phi_0$, the field oscillates with a slowly decreasing
amplitude until it approaches $\phi_0$. But the energy density cannot
decrease too much because at the moment when $\rho = V(\phi) +
\dot\phi^2/2$ vanishes, the Hubble constant vanishes too, so that
$\dot a = 0$. Then the universe begins to
collapse, $\dot a <0$, and the amplitude of the oscillations begins to
grow. Eventually this growth becomes so fast that the field stops
oscillating and moves towards $\phi = \pm \infty$.

The best way to understand this effect is to examine what happens
during the critical oscillation when the sign of $\dot a$ changes.  We
will study this process analytically, making some simplifying
approximations.

First of all, we will assume that the field $\phi$ begins this
oscillation at $t=0$ moving with zero initial velocity from a point
$\phi_1\approx \phi_0$ such that $0< \Delta\phi = \phi_1 -\phi_0 \ll
\phi_0$. The initial energy density of the field is $\Delta V =
V(\phi_1) = {m^2\over 2}(\phi_1^2 -\phi_0^2) \ll |V(0)|$.  We will try
 to evaluate the turning point moment $t_c$ where $\dot a= 0$
(i.e. $H=0$).

Let us consider the series expansion of the Hubble parameter around the
beginning of this process
\begin{equation}\label{hubble}
H(t) \approx H_1+  \dot H_1  t +{ 1 \over 2}   \ddot H_1  t^2 +
{1 \over 3!}\,     {}^{{}^{{}^{\bf \centerdot \centerdot \centerdot }}}\hskip -0.37cm H_1   t^3  + ... \ ,
\end{equation}
where $ H_1$ and its derivatives are taken at $t=0$.  The reason to
include the terms up to $t^3$ in this series is the following. From
the relation $\dot H=- {1 \over 2}\dot \phi^2$ we find that for
vanishing initial velocity $\dot \phi_1=0$ one has $ \dot H_1 = \ddot
H_1=0$. The first nonvanishing coefficient ${}^{{}^{{}^{\bf \centerdot
\centerdot \centerdot }}}\hskip -0.37cm H_1 \approx - \ddot \phi^2
\approx -(V'(\phi_1))^2= - m^4 \phi_1^2$ is negative. Note that $H_1 =
\sqrt{V(\phi_1)/3}= \sqrt{\Delta V/3}$. This means that at the moment
\begin{eqnarray}\label{turning}
t_c \approx \left ({12\, V(\phi_1)\over (V'(\phi_1))^4}\right)^{1/6} = m^{-1}  \left ({12\, \Delta V\over m^2 \phi_0^4}\right)^{1/6}
\end{eqnarray}
the Hubble parameter vanishes. Note that the first part of this
equation is pretty general, whereas the second one is specific to
quadratic potentials.

At the turning point
\begin{eqnarray}\label{turning1}
\phi_c \approx \phi_0 - \left ({3\Delta V\over 2 m^2 \phi_0}\right)^{1/3}  \ .
\end{eqnarray}
These results imply that the turn occurs during the first oscillation
starting at $\phi_1$ if $\Delta V \lesssim m^2 \phi_0^4$,
i.e. $\phi_1^2-\phi_0^2 \lesssim \phi_0^4$. In the most interesting
case $\Delta V \ll m^2 \phi_0^4$ the turn occurs in the immediate
vicinity of the point $\phi_0$ where the potential becomes negative.

To study the subsequent evolution of $\phi(t)$ and $a(t)$, let us
assume that the scale factor $a$ during the first oscillation does not
change much. This is a reasonable assumption since $\dot a= 0$ at the
turning point. We will therefore take $a = 1$ during this oscillation,
and $\phi(t) = \phi_1~ \cos mt$. The potential energy density of the
field is
\begin{eqnarray}\label{poten}
V(\phi) = \Delta V- {m^2 \phi_1^2\over 2} ~\sin^2 mt \ 
\end{eqnarray}
and the acceleration of the universe is given by
\begin{eqnarray}\label{acceler}
\ddot a \approx {\ddot a\over a} = {V-\dot\phi^2\over 3} ={\Delta
V\over 3} - {m^2 \phi_1^2\over 2} ~\sin^2 mt \ .
\end{eqnarray}
Taking into account that initially $\dot a = a \sqrt {\Delta V/3}
\approx \sqrt {\Delta V/3}$, this yields
\begin{eqnarray}\label{acceler2}
\dot a \approx {\Delta V\over 3} t - {m^2 \phi_1^2\over 4}t + {m
\phi_1^2\over 8} ~\sin 2mt + \sqrt {\Delta V/3}\ .
\end{eqnarray}
By integrating this relation from $t = 0$ to $t = \pi/m$, i.e. during
one half of an oscillation, one finds that the condition $a \approx 1$
implies then that $\phi_1 \approx \phi_0 \ll 1$, i.e $\phi_0 \ll M_p$.

Now we are going to find how the energy density $\rho$ of the field $\phi$ changes during the time $\pi/m$ when the field $\phi$ moves from $\phi_1$ to $-\phi_1$. In order to do it, we will represent 
the scalar field equation $\ddot\phi  +3H\dot\phi = -V'(\phi)$  in the form 
\begin{eqnarray}\label{energy}
\dot \rho = {d (V+\dot\phi^2/2)\over dt} = -3H\dot\phi^2   \ .
\end{eqnarray}
Thus in order to find the total change of the energy density of the scalar field during some time one should integrate $-3H\dot\phi^2$:
\begin{eqnarray}\label{energychange}
\Delta \rho = \Delta (V+\dot\phi^2/2) = -3\int\limits_{t_0}^{t} H\dot\phi^2  dt \ .
\end{eqnarray}
Using this equation, one can find the change of the energy density of the field $\phi$  during the time $\pi/m$ when the field $\phi$ moves from $\phi_1$ to $-\phi_1$:
\begin{eqnarray}\label{engrowth}
\Delta \rho_- = {3\pi^2\over 16} m^2\phi_1^4 - {\pi \sqrt {3\Delta V}\over 2}m\phi_1^2 - {\pi^2\over 4} \Delta V \phi_1^2 \ .
\end{eqnarray}
In the most interesting case  $\phi_1 \approx \phi_0$, one can neglect the last term in this equation and replace $\phi_1$ by $\phi_0$:  
\begin{eqnarray}\label{engrowthsimple}
\Delta \rho_- = {3\pi^2\over 16} m^2\phi_0^4 - {\pi \sqrt {3\Delta V}\over 2}m\phi_0^2  \ .
\end{eqnarray}
Thus, if the initial kinetic energy  of the field is equal to zero at the beginning of the oscillation at $\phi = \phi_1$, at the moment when the field $\phi$ will reach the point $-\phi_1$ its kinetic energy will  be positive,
\begin{eqnarray}\label{engrowthsimple2}
 {\dot\phi^2\over 2}  = \Delta \rho_- =  {3\pi^2\over 16} m^2\phi_0^4 - {\pi \sqrt {3\Delta V}\over 2}m\phi_0^2  \ .
\end{eqnarray}

Note that for $\Delta V \ll m^2 \phi_0^4$ the last term is much smaller than the first one, so one finds, in the first approximation, that the field $\phi$ coming to the point $-\phi_1$ acquires kinetic energy 
\begin{eqnarray}\label{engrowthsimple2ii}
 {\dot\phi^2\over 2}  = \Delta \rho_- \approx  {3\pi^2\over 16} m^2\phi_0^4  = {3\pi^2 \phi_0^2\over 8}~ V_0  \ll V_0\ ,
\end{eqnarray}
and velocity
\begin{eqnarray}\label{engrowthsimple3}
\dot \phi \approx \sqrt{3\pi^2\over 8} m \phi_0^2    \ .
\end{eqnarray}
This velocity continues to grow during subsequent oscillations and
eventually the scalar field $\phi$  and the scale factor $a$ blow 
up,  as shown in Fig. \ref{qfig2}.

So far we have studied an expanding  universe that stops its expansion
and collapses. But what if it was collapsing at the beginning of the
oscillation? Suppose the scalar field was moving very slowly until it
reached the point $\phi_1$.   Then it started falling down, just as in
the case considered above. However, this time we will assume that the
universe  was not expanding but collapsing. This corresponds to the
choice $\dot a = -\sqrt{\Delta V/3}$ at the beginning of the process.

In this case the universe will continue collapsing with ever growing
speed. The evolution of the field $\phi$ can be studied by the same
methods as the ones used above. The main difference will be that the
field $\phi$ passing through the point $\phi = -\phi_1 \approx \phi_0$
will have kinetic energy  

\begin{eqnarray}\label{engrowthsimple2a}
{\dot\phi^2\over 2}  = \Delta \rho_+ = {3\pi^2\over 16} m^2\phi_0^4 + {\pi \sqrt {3\Delta V}\over 2}m\phi_0^2  \ .
\end{eqnarray} 

The kinetic energy of the field $\phi$ at $\phi = -\phi_0$ differs from that at $\phi = -\phi_1$ by $\Delta V$. However, for $\Delta V \ll m^2 \phi_0^4$ this difference is much smaller than each of the terms in Eqs. (\ref{engrowthsimple}), (\ref{engrowthsimple2a}). Thus these two equations with the above-mentioned accuracy give the kinetic energy of the field $\phi$ not only at $\phi = -\phi_1$ but also at $\phi = -\phi_0$. 

This discussion, as well as the  difference between $\Delta \rho_-$ and $\Delta \rho_+$, will play an important role in our investigation of the cyclic universe scenario \cite{Cyclic}. As we will see, the cyclic regime is possible only if the field $\phi$, after bouncing from the singularity, approaches the point $-\phi_0$ with energy density greater than $\Delta \rho_+$, which in its turn is greater than $\Delta \rho_-$, which is the energy of this field at the point $-\phi_0$ on its way towards the singularity. Thus one needs this field to bounce from the singularity with an increased energy, and one should check that the possible source of this additional energy does not create problems for the scenario.  
In fact, we will see that with an account taken of particle production, the required energy increase can be much greater 
than the difference between $\Delta \rho_+ $ and $\Delta \rho_-$. 

\section{\label{Other}Other models with  $V(\phi) <0 $}

Until now we have studied only one simple model with a quadratic
potential. However, many features of models with negative potentials
are model-independent. Consider, for example, the model with the
``inverted potential'' $V(\phi) = V_0 - {m^2\phi^2\over 2}$ with $V_0
>0$. This is the simplest example of a potential unbounded from
below. The evolution of the scalar field and scale factor in this
model is shown in Fig. \ref{conjunto4}.  As we see, in the beginning
the universe experiences a stage of inflation when the scalar field
slowly rolls from the top of the effective potential. (We considered a
model with $V_0 \gg m^2$.)  Later on, inflation ends and the speed of
the field increases. If one neglects the effects of the expansion of
the universe, at large $\phi$ one has $\dot\phi^2 =
2(V_0-V(\phi))$. Therefore
\begin{equation}
{\ddot a\over a} = {1\over 3} (V(\phi)- \dot\phi^2 ) = V(\phi)
-{2\over 3} V_0 \ .
\end{equation}
At large $\phi$ the universe starts moving with ever growing negative
acceleration. If one takes into account the expansion of the universe,
$\dot\phi^2$ becomes even smaller, and the deceleration is even
greater.  As a result, the expansion slows down and the universe
starts contracting.  At this stage the ``friction term'' $3H\dot\phi$
in the equation of motion of the scalar field becomes negative, which
causes the field $\phi$ to grow and leads to a rapid collapse of the
universe.

\FIGURE[!h]{\epsfig{file=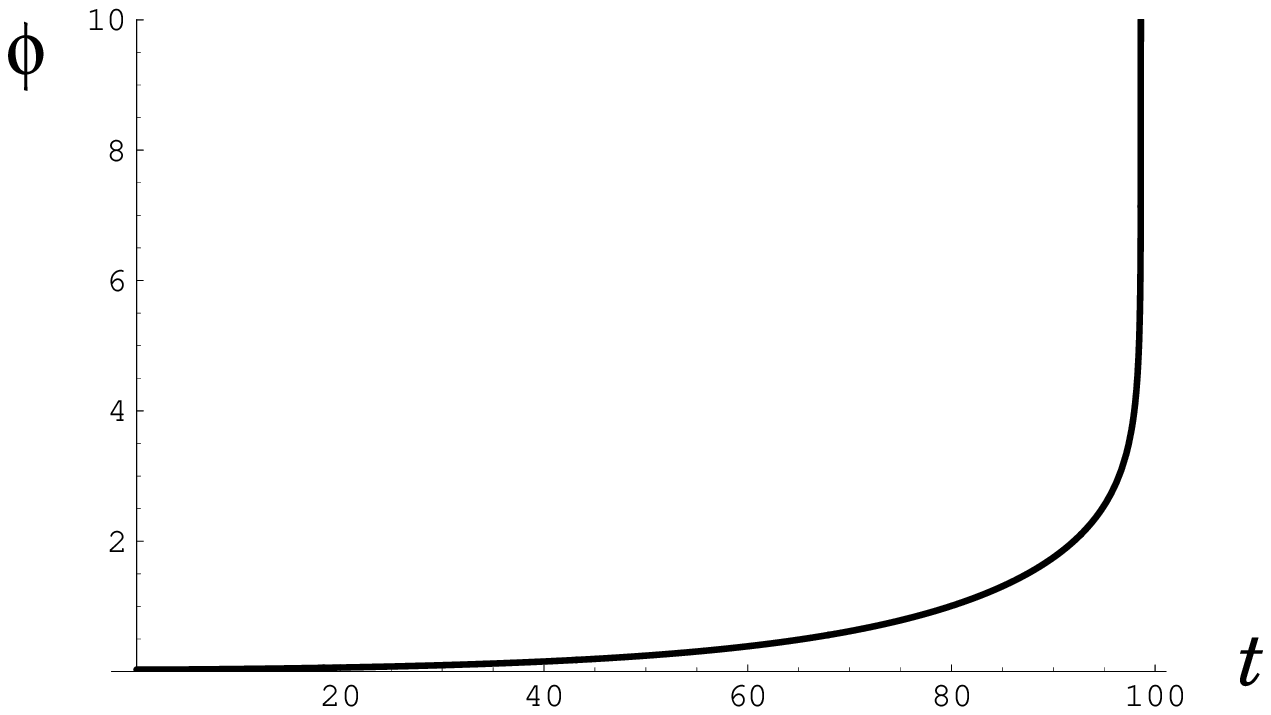,width=.47\columnwidth} \qquad
\epsfig{file=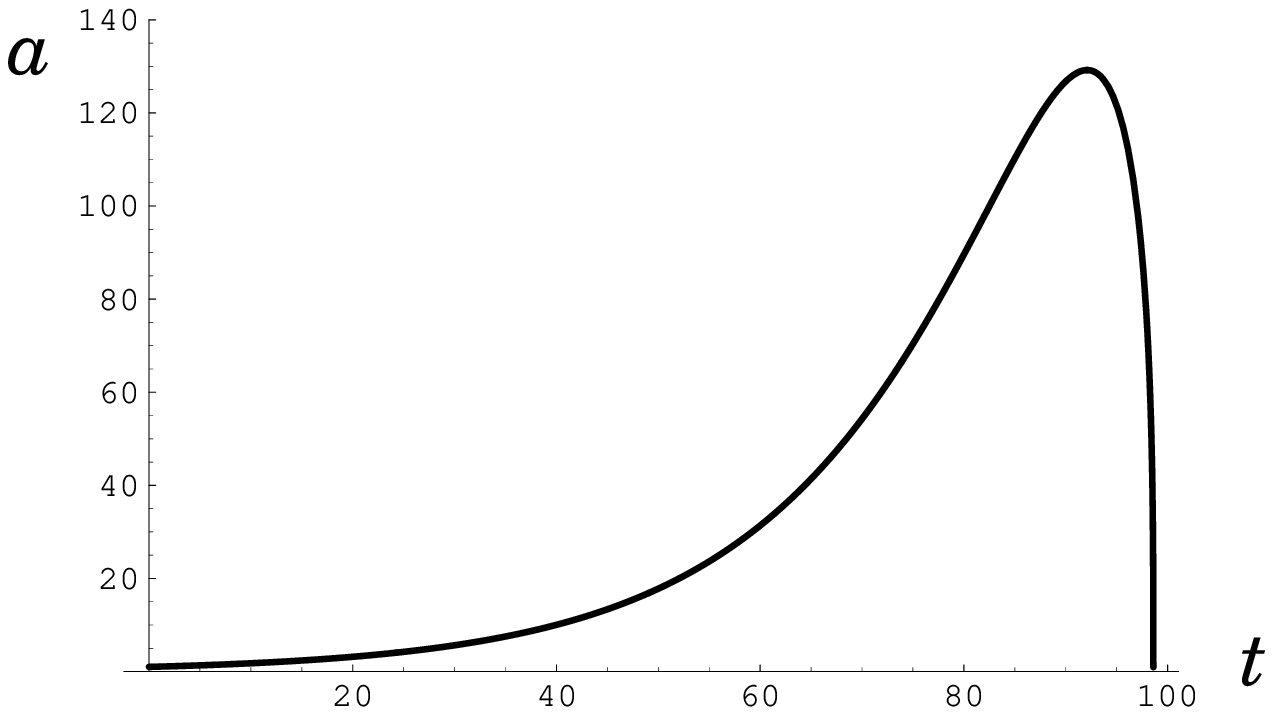,width=.47\columnwidth} \caption{Evolution of the
scalar field and scale factor in the model $V(\phi) = V_0 -
{m^2\phi^2\over 2}$.} \label{conjunto4}}

Another example is the standard potential used for the description of
spontaneous symmetry breaking, with the addition of a negative
cosmological constant $V_0 <0$:
\begin{eqnarray} \label{SSB}
V(\phi) = {\lambda\over 4} (\phi^2-v^2)^2 +V_0 =
-{1\over 2} m^2\phi^2 + {m^2 \over 4 v^2} \phi^4 +{1\over 4} m^2 v^2
+V_0 \ .
\end{eqnarray}
Here $m^2 \equiv \lambda v^2$ and the point $\phi = v$ corresponds to
the minimum of $V(\phi)$ with symmetry breaking. The potential
$V(\phi)$ becomes equal to $V_0 <0$ in the minimum of $V(\phi)$ at
$\phi = v$.  As we see in Fig. \ref{conjunto4a}, the scalar field in
this case experiences a stage of oscillations near the minimum of the
effective potential with $V(\phi)= V_0 <0$, but then it jumps off the
minimum and blows up because of the ``negative friction'' in the
collapsing universe. For most model parameters and initial conditions,
if the field originally moves towards the minimum with $\phi = +v$ it
will blow up in the direction $\phi \to -\infty$ and {\it vice
versa}. The reason is that at the initial stages of the development of
the instability the field $\phi$ is most efficiently accelerated by
the negative friction if for a while it moves in a relatively flat
direction, i.e.  from one minimum to another, instead of directly
moving upwards \cite{Fastroll}.

\

\FIGURE[!h]{\epsfig{file=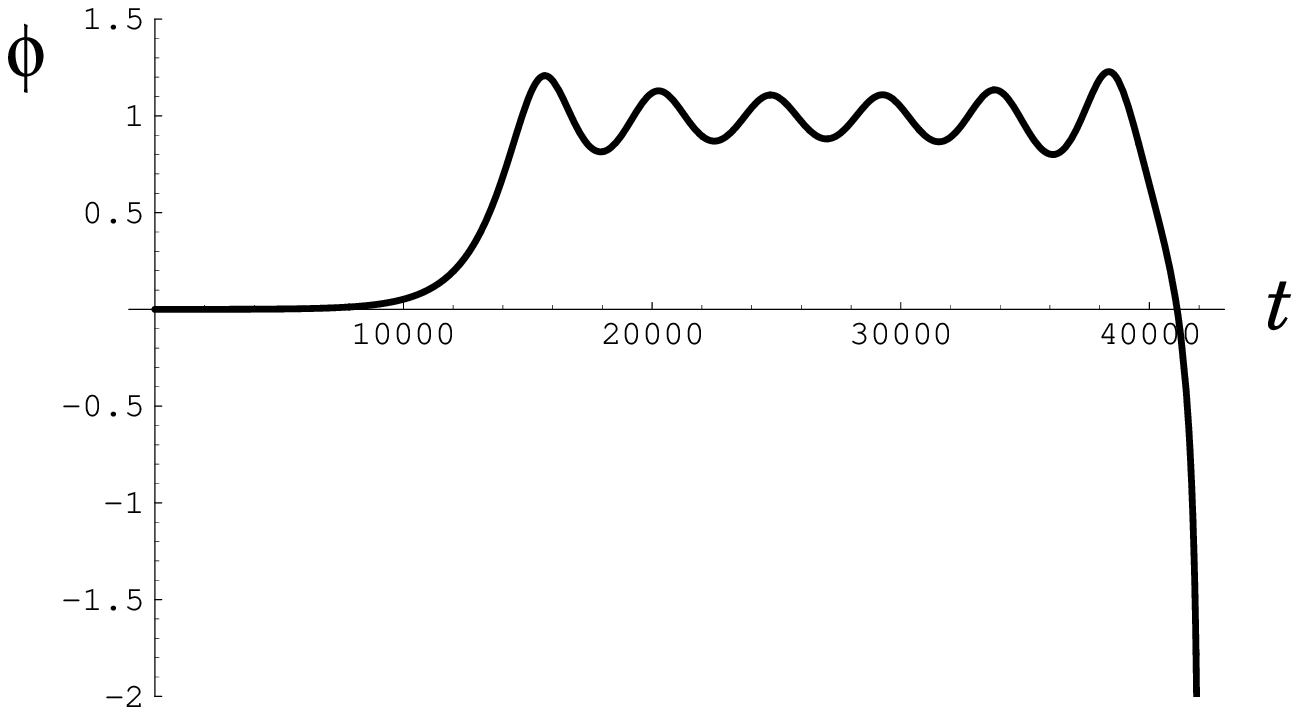,width=.47\columnwidth} \qquad
\epsfig{file=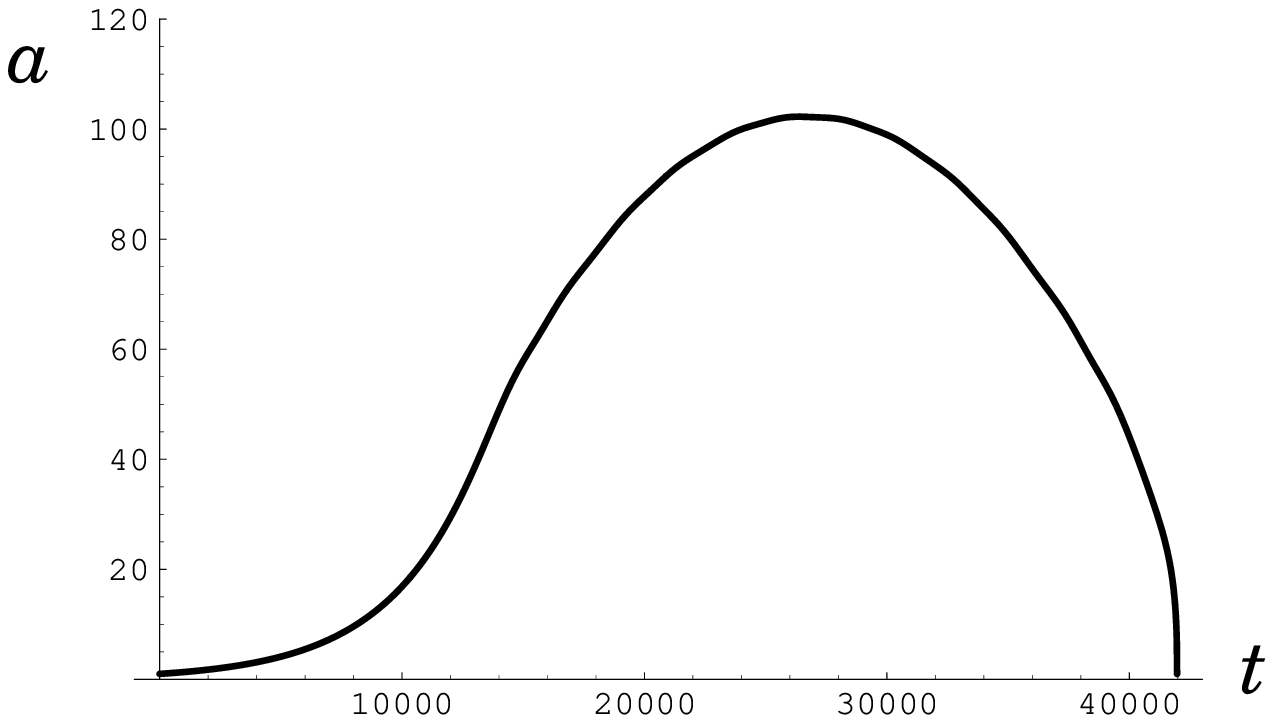,width=.47\columnwidth} \caption{Evolution of the
scalar field and the scale factor in the model $V(\phi) =
{\lambda\over 4} (\phi^2-v^2)^2 + V_0$, with $V_0 <0$.}
\label{conjunto4a}}

When the field accelerates enough it enters the regime $\dot\phi^2 \gg
V(\phi)$ and continues growing with a speed practically independent
of $V(\phi)$: $\phi \sim \ln t$, $\dot\phi \sim t^{-1}$, see
Eq. (\ref{1.7.25a}). So for all potentials $V(\phi)$ growing at large
$\phi$ no faster than some power of $\phi$ one has $\dot\phi^2/2$
growing much faster than $V(\phi)$ (a power law singularity versus a
logarithmic singularity). This means that one can indeed neglect
$V(\phi)$ in the investigation of the singularity, virtually
independently of the choice of the potential. Thus we see that from
the point of view of the singular behavior of the field $\phi(t)$ and
the scale factor, {\it potentials having a global minimum with
$V(\phi) <0$ are as dangerous as potentials unbounded from below.}

Since a small modification of the potential (shifting the minimum of
$V(\phi)$ towards $V(\phi) < 0$) may lead to a change of regime from
expansion to contraction, one may wonder whether some other
modification of $V(\phi)$ can switch the regime of contraction back to
expansion? The answer follows from the equation $\dot H = - \frac{1}{2}
(\rho + p)$. This equation implies that $\dot H \leq 0$ because $\rho
+ p \geq 0$ in accordance with the null energy condition. This means,
in particular, that if the universe switches from expansion to
contraction, it cannot later return to the regime of expansion. The
only possible exception would be if the universe were to pass through
a stage of super-Planckian density in which the Einstein equations
were invalid.

Even though many properties of the theories with negative potentials
are model-independent, the topology of their phase portraits depends
on the choice of the potential $V(\phi)$. For example, the
hypersurface representing a flat universe in the theory $V(\phi) = V_0
- {m^2\phi^2\over 2}$ is given by the constraint equation
\begin{equation} \label{dual}
\dot\phi^2  = 6H^2 + m^2 \phi^2    - 2 V_0\ .
\end{equation}
This equation describes a hyperboloid just like the flat universe
hypersurface of the theory $V(\phi) = V_0 + {m^2\phi^2\over 2}$. In
this case, however, the axis of the hyperboloid is in the $\dot{\phi}$
direction rather than the $H$ direction. Moreover, the hyperboloid for
this model has two sheets for $V_0<0$ and one sheet for $V_0>0$, which
is the reverse of the situation for $V(\phi) = V_0 + {m^2\phi^2\over
2}$. The different orientation of the hyperboloid means, for example,
that for the theory unbounded from below all trajectories end in a Big
Crunch singularity, regardless of the signs of $V_0$ and $k$.

\section{\label{singularity}Approach to the Singularity, Quantum
Corrections, and Particle Production}
 
Talking about the dynamics of the cosmological scalar field, until now
we have remained in the realm of classical physics. We ignored
possible quantum effects, and in particular the effects of particle
production. These effects may lead to some important qualitative
changes of the phase portraits, however, especially near the
singularity.

First of all, near the singularity one may need to take into account
quantum corrections to the effective action of general
relativity. Even ignoring possible effects related to brane cosmology
or M-theory, one may need to add to the effective action terms
proportional to $R^2$, $R_{\mu\nu} R^{\mu\nu}$, etc.

An important example of such a theory is given by a combination of
scalar field theory and the Starobinsky model, where the effective
Lagrangian has additional terms $\sim R^2$ \cite{KLS}. Whereas this
addition is not very significant at low energies, it completely
changes the behavior of the theory near the singularity.

For example, in the absence of this term the generic regime for a
scalar field approaching the singularity is $\dot\phi^2/2 \gg
V(\phi)$, which corresponds to the equation of state $p =\rho$. This
regime was recently discussed in \cite{BankFish} in the context of
string cosmology. As we have seen, in this case $a \sim t^{1/3}$,
$\phi \sim \ln t$.

However, if one adds the term $R^2$, the most general regime for
theories where the potential is not too steep becomes quite different:
$a \sim t^{1/2}$, $\phi \sim t^{-1/2}$ \cite{KLS}.

It is even more important to consider the effects of particle
production. If one ignores quantum effects, one typically finds the
curvature $R \sim t^{-2}$ in a collapsing universe. Scalar particles
minimally coupled to gravity, as well as gravitons and helicity 1/2
gravitinos \cite{Kallosh:2000ve}, are not conformally invariant; their
frequencies thus experience rapid nonadiabatic changes induced by the
changing curvature. These changes lead to particle production due to
nonadiabaticity with typical momenta $k^2 \sim R \sim t^{-2}$.  The
total energy-momentum tensor of such particles produced at a time $t$
after (or before) the singularity is $T_{\mu\nu} \sim O(k^4) \sim R^2
\sim t^{-4}$ \cite{ford,grib}. Comparing the density of
produced particles with the classical matter or radiation density of
the universe $\rho \sim t^{-2}$, one finds that the density of created
particles produced at the Planck time $t \sim 1$ is of the same order
as the total energy density in the universe.

The main point of this
discussion is that particle production near a cosmological singularity
can be extremely efficient. Generically one expects that when the
universe emerges from  or approaches a singularity and its density
is close to the Planck density, the density of produced particles
should be comparable to the total energy density of the universe.
 
This is a pretty general conclusion. For example, in brane cosmology a
similar effect of particle production may occur even though $R = 0$ in
4D. Indeed, the change of distance between branes leads to a
nonadiabatic change of the spectrum of Kaluza-Klein modes and thus to
particle production; one may call it a time-dependent Casimir
effect. Note that this effect exists even in theories with unbroken
supersymmetry \cite{Maroto}.  

This observation has many implications. In particular, one can no
longer expect that matter (or a scalar field) has the equation of
state $p = \rho$ near the singularity. Even if the universe around the
Planck time was dominated by matter with $p = \rho$, the creation of
particles would immediately change the situation. And even if the
density of created particles initially was somewhat smaller than the
energy density of matter with $p = \rho$, this situation would rapidly
change. The density of the component of matter with $p = \rho$
decreases as $a^{-6}$, whereas the energy density of radiation and
nonrelativistic particles decrease as $a^{-4}$ and $a^{-3}$
respectively. Therefore the energy density of such particles soon
becomes greater than the energy density of the matter component with
$p = \rho$. Once this happens the scalar field immediately freezes. It
loses its initial kinetic energy and begins moving very slowly. As we
already discussed, this provides perfect initial conditions for
inflation. This result also has important implications for the cyclic
universe scenario \cite{Cyclic}.

\section{\label{Cycles}Cyclic universe}

\subsection{\label{epicyclic}The basic scenario}

Until now we have studied the evolution of the universe and classified
new possibilities that appear in scalar theories with negative
potentials. This problem is very interesting. Its investigation has
already brought us to an important realization: We cannot live in anti
de Sitter space dominated by a negative cosmological constant, not
because the negative cosmological constant is forbidden, but because a
universe dominated by negative vacuum energy cannot appear after a
long stage of inflation \cite{Banks:1995dt,N8,Fastroll}. Another interesting
realization is that the available observational data can tell us
nothing about the future of the universe: we may live in a stage of a
nearly constant de Sitter-like inflationary acceleration, but it may
end with a global collapse
\cite{Krauss:1999br,Kaloper:1999tt,Starobinsky:2000yw,N8,Fastroll}.

A common feature of cosmological evolution in models with negative
potentials is that it begins in a singularity {\it and ends in a
singularity}, even if the universe is not closed. This was not the
case for the theories with $V(\phi) > 0$, where the universe may
continue expanding forever and never end in a singularity even if it
is closed.

This naturally brought back old speculations about the oscillating, or
cyclic, evolution of the universe, see e.g. \cite{Tolman-1931} -
\cite{PBB}, \cite{Cyclic}. The universe may be created in a
singularity, then collapse and re-emerge again.

There is a certain intellectual attractiveness in this idea. However,
during the last 20 years this idea has lost some of its initial
appeal. Indeed, if there was a stage of inflation after the
singularity, then the initial conditions producing our universe are
nearly irrelevant for the investigation of the formation of
large-scale structure in the observable part of the
universe. Moreover, inflation in many of its simplest versions is
eternal \cite{VilEt,Eternal}. This fact may not solve the singularity
problem \cite{Borde},
but it puts the origin of our part of the universe
indefinitely far away in the past \cite{LLM}.

Recently Steinhardt and Turok proposed a version of inflationary
theory where the stage of inflation occurs {\it after} formation of
the large scale structure of the universe and perturbations
responsible for the formation of the structure of the universe are
produced {\it before} the singularity, during the previous cycle of
the universe evolution \cite{Cyclic}. In this scenario inflation does
not protect us from all uncertainties associated with the physical
processes  occurring around the big bang. On the contrary, in order
to describe our universe in this scenario one must know exactly what
happens with small perturbations of the metric when they pass through
the singularity.

The cyclic scenario \cite{Cyclic} is a modified version of the
ekpyrotic scenario \cite{KOST}. It is based on the idea that we live
on one of two branes whose separation can be parametrized by a scalar
field $\phi$.  It is assumed that one can describe the brane
interaction by an effective 4D theory with the effective potential
$V(\phi)$ having a minimum at $V(\phi) <0$. In the original version of
the ekpyrotic scenario it was assumed that $V(\phi)$ is always
negative, but it vanishes at $\phi = 0$ and at $\phi \to \infty$.  It
was claimed that one of the main advantages of the ekpyrotic scenario
was the absence of a cosmological singularity and the possibility to
solve the major cosmological problems without the help of inflation,
which was called ``superluminal expansion.''

However, later it was found that it is difficult to solve the
cosmological problems in the ekpyrotic scenario without using
inflation \cite{pyrotech}.  Moreover, perturbations of the field
$\phi$ that could be responsible for large scale structure formation
in this scenario are generated due to tachyonic instability 
\cite{FKL} at the
time when $V(\phi)$ was supposed to be smaller than $-10^{-50}$ in
Planck units. Therefore it is difficult to avoid inflation in this
model: Even a miniscule positive contribution to $V(\phi)$ of the
order $10^{-50}$ would lead to a stage of exponential expansion of the
universe at large $\phi$ \cite{pyrotech}. Also, in \cite{KKLT} it was
shown that in the context of the effective 4D theory used in
\cite{KOST} the universe can only collapse. This means that the
ekpyrotic scenario suffers from the cosmological singularity
problem. This problem has been analysed in \cite{Seiberg}, but so far
it remains unresolved.

In the cyclic universe scenario the authors assume, in accordance with
the suggestion of Ref. \cite{pyrotech}, that $V(\phi)$ is positive at
large $\phi$, and therefore the universe experiences a stage of
inflation. This stage provides the solution to the major cosmological
problems. However, it is assumed that this is an extremely low-scale
inflation associated with the present stage of acceleration of the
universe in a state with $V(\phi) \sim 10^{-120}$. Inflationary
perturbations produced at this stage have wavelengths comparable to
the present size of the horizon, so they cannot be responsible for
galaxy formation.

\begin{figure}[h!]
\leavevmode
\centering \epsfysize=4cm
\begin{tabular}{ccc}
\includegraphics[scale=0.9]{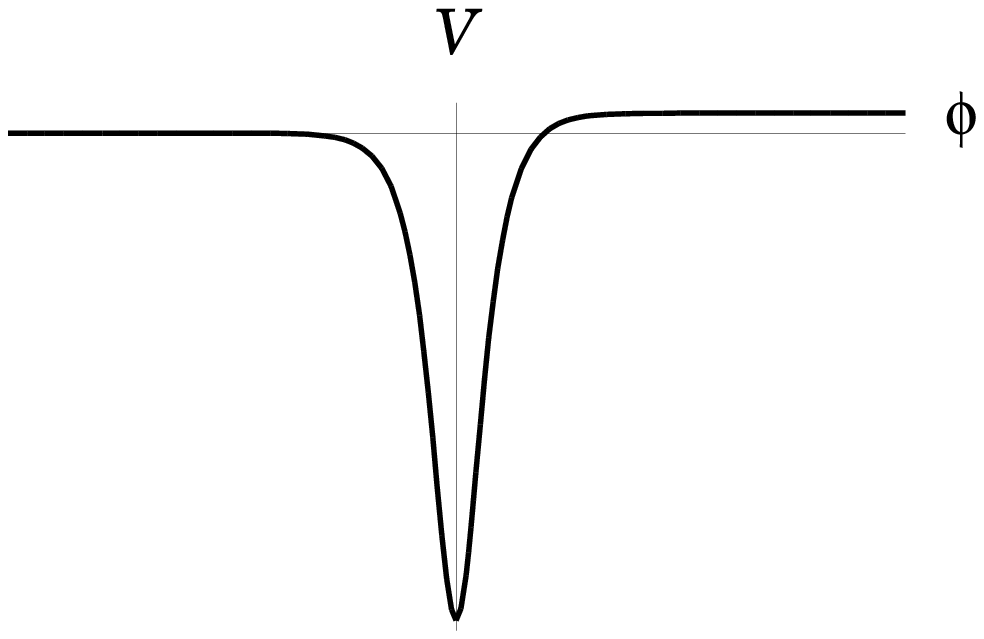} & \qquad 
\end{tabular}
\caption{Scalar field potential in the cyclic scenario. The minimum of
the potential may occur at any value of $\phi$; in this section for
simplicity we will assume that it occurs at $\phi = 0$; we will
consider a more general situation later. }
\label{Cyclic}
\end{figure}

Therefore it is assumed that the desired perturbations of the scalar
field are produced after inflation, by the same tachyonic mechanism as
in the ekpyrotic scenario \cite{FKL,KOST,pyrotech}. The effective
potential of the scalar field in the cyclic scenario has the shape
shown in Fig. \ref{Cyclic}. Inflation occurs at large $\phi$. Once the
field rolls down to the region where $V(\phi)<0$, the universe begins
to collapse. At that time perturbations of the scalar field are
generated. The speed of the field in a collapsing universe grows. It
reaches the plateau at $\phi \to -\infty$ where, according to
\cite{Cyclic}, the potential vanishes. The universe enters the regime
where its energy density is dominated by the kinetic energy of the
scalar field, and it evolves towards the singularity in accordance
with Eqs. (\ref{akin}), (\ref{1.7.25a}).

Usually, this would be considered the end of the evolution of the
universe. However, in the cyclic scenario it is assumed that the
universe goes through the singularity and re-appears again. When
it appears, in the first approximation it looks exactly as it was
before, and the scalar field moves back exactly by the same trajectory
by which it reached the singularity \cite{Seiberg}.

This is not a desirable cyclic regime. Therefore it is assumed in
\cite{Cyclic} that the value of kinetic energy of the field $\phi$
{\it increases} after the bounce from the singularity. This increase
is supposed to appear as a result of particle production at the moment
of the brane collision (even though one could argue that usually
particle production leads to an opposite effect). If the increase of
the kinetic energy is large enough, the field $\phi$ rapidly rolls
over the minimum of $V(\phi)$ in a state with a positive total energy
density, and continues its motion at $\phi > 0$.  The kinetic energy
of the field decreases faster than the energy of matter produced at
the singularity.  At some moment the energy of matter begins to
dominate. Eventually (a few billion years after the big bang) galaxies
form. Then the energy density of ordinary matter becomes smaller than
$V(\phi)$ and the present stage of inflation (acceleration of the
universe) starts again.

As we see, this version of the ekpyrotic scenario is not an
alternative to inflation anymore. Rather it is a very specific version
of inflationary theory. The major cosmological problems are supposed
to be solved due to exponential expansion in a vacuum-like state, even
though the mechanism of production of density perturbations in this
scenario is non-standard. Let us remember that Guth's first paper on
inflation \cite{Old} was greeted with so much enthusiasm precisely
because it proposed a solution to the homogeneity, isotropy, flatness
and horizon problems, even though it didn't address the formation of
large scale structure. The Starobinsky model that was proposed a year
earlier \cite{Starobinsky:te} could account for large scale structure
and the observed CMB anisotropy \cite{pert}, but it did not
attract as much attention because it did not emphasize the possibility
of solving these initial condition problems.

In fact, the stage of acceleration of the universe in the cyclic model
is {\it eternal inflation}. Indeed, the main criterion for the process
of self-reproduction of the inflationary universe to occur is that the
amplitude of inflationary perturbations $\delta\phi \sim H \sim \sqrt
V$ should be greater than the change $\Delta \phi$ of the classical
value of the field $\phi$ during the time $H^{-1}$: ~$\Delta \phi \sim
V'/V$ \cite{VilEt,Eternal,LLM}. For the potential $V(\phi)$ used in
the cyclic model one has $\delta\phi = const$ in the limit $\phi \to
\infty$, whereas $\Delta\phi \to 0$ in this limit. Thus the universe
at large $\phi$ enters the stage of eternal self-reproduction, quite
independently of the possibility to go through the singularity and
re-appear again. In other words, the universe in the cyclic scenario
is not just a chain of eternal repetition, but a growing
self-reproducing inflationary fractal of the type discussed in
\cite{VilEt,Eternal,LLM}.\footnote{It is remarkable that quantum effects and the mechanism of self-reproduction may work even at the present  stage when the wavelength of inflationary fluctuations 
is greater than the size of the observable part of the universe and the square of their amplitude is as small as $10^{-120}$ in Planck units. The reason why it may work is that the curvature of the effective potential at large $\phi$ is even much smaller.}

One may wonder, however, whether this version of inflationary theory
is good enough to solve all major cosmological problems. Indeed,
inflation in this scenario may occur only at a density 120 orders of
magnitude smaller than the Planck density. If, for example, one
considers a closed universe filled with matter and a scalar field with
the potential used in the cyclic model, it will typically collapse
within the Planck time $t \sim 1$, so it will not survive until the
beginning of inflation in this model at $t \sim 10^{60}$.  For
consistency of this scenario, the overall size of the universe at the
Planck time must be greater than $l \sim 10^{30}$ in Planck units,
which constitutes the usual flatness problem. The total entropy of a hot universe that may survive until
the beginning of inflation at $V \sim 10^{-120}$ should be greater
than $10^{90}$, which is the entropy problem \cite{book}.  An estimate
of the probability of quantum creation of such a universe ``from
nothing'' gives $P \sim e^{-|S|} \sim \exp\left(-{ 24\pi\over V
}\right) \sim e^{-120}$ \cite{Creation}.

There are some other unsolved problems related to this theory, such as
the origin of the potential $V(\phi)$ \cite{pyrotech} and the 5D
description of the process of brane motion and collision
\cite{KKLT,Rasanen:2001hf}. In particular, the cyclic scenario assumes
that the distance between the branes is not stabilized. Thus one would
need to find some other mechanism that would ensure that the effective
gravitational constant, as well as other parameters depending on the
field $\phi$ (i.e. on the brane separation), does not change in time
too fast. This is one of the reasons why it is usually assumed that
the branes in Ho\u{r}ava-Witten theory must be stabilized.

We will not discuss these problems here. Instead of that, we will
concentrate on the phenomenological description of possible cycles
using the effective 4D description of this scenario. This will allow
us to find out whether the cyclic regime is indeed a natural feature
of the scenario proposed in \cite{Cyclic}.

For the remainder of this section we will analyse this scenario using
the tools developed in the earlier sections of the paper. In Section
\ref{seccycport} we will describe the phase portrait of the cyclic
scenario. In Section \ref{moving} we will consider the conditions that
must be satisfied at the bounce in order for the cyclic regime to
occur. In Section \ref{afterbounce} we will analyse the motion of the
field as it returns from the singularity and show that the conditions
described in Section \ref{moving} are difficult to realize
self-consistently without invoking super-Planckian potentials, even in
the vicinity of the minimum. Following the authors of \cite{Cyclic} we
will consider such super-Planckian potentials in Section
\ref{superplanck}. Aside from the problem of applying the effective 4D
theory at such high energies, we will find that there are still other
problems in such realizations of the scenario. In Sections
\ref{bicycling} and \ref{cyclicinflation} we will propose some
modifications of the cyclic scenario that may resolve some of the
problems raised here.

\subsection{\label{seccycport} Phase portrait of the cyclic universe} 

The phase space of  the cyclic scenario is the usual 3d space $(\phi,\dot\phi,H)$.
If one does not take into account matter and radiation, the phase portrait of the  scenario forms a 2d surface in 3d space. It  is shown in Fig. \ref{cycport}  without the Poincar\'{e} mapping. (If one adds  radiation, the flow of trajectories becomes
three dimensional.) The trajectories corresponding to different
initial values of $\phi$ and $\dot\phi$ start at large $H$, i.e. in
the upper part of Fig. \ref{cycport}. The trajectories beginning at
large positive $\phi$ reach the (red) separatrix going from the point
S1 to the point A. Its upper part ($H > 0$) corresponds to inflation.  These
trajectories follow the separatrix towards the throat of the phase
portrait at $H = 0$, and then all of them move towards the
singularity.  The trajectories beginning at large negative $\phi$ fall
from the singularity at large positive $H$ to the singularity at large
negative $H$ without entering the stage of inflation.

\FIGURE[!h]{\epsfig{file=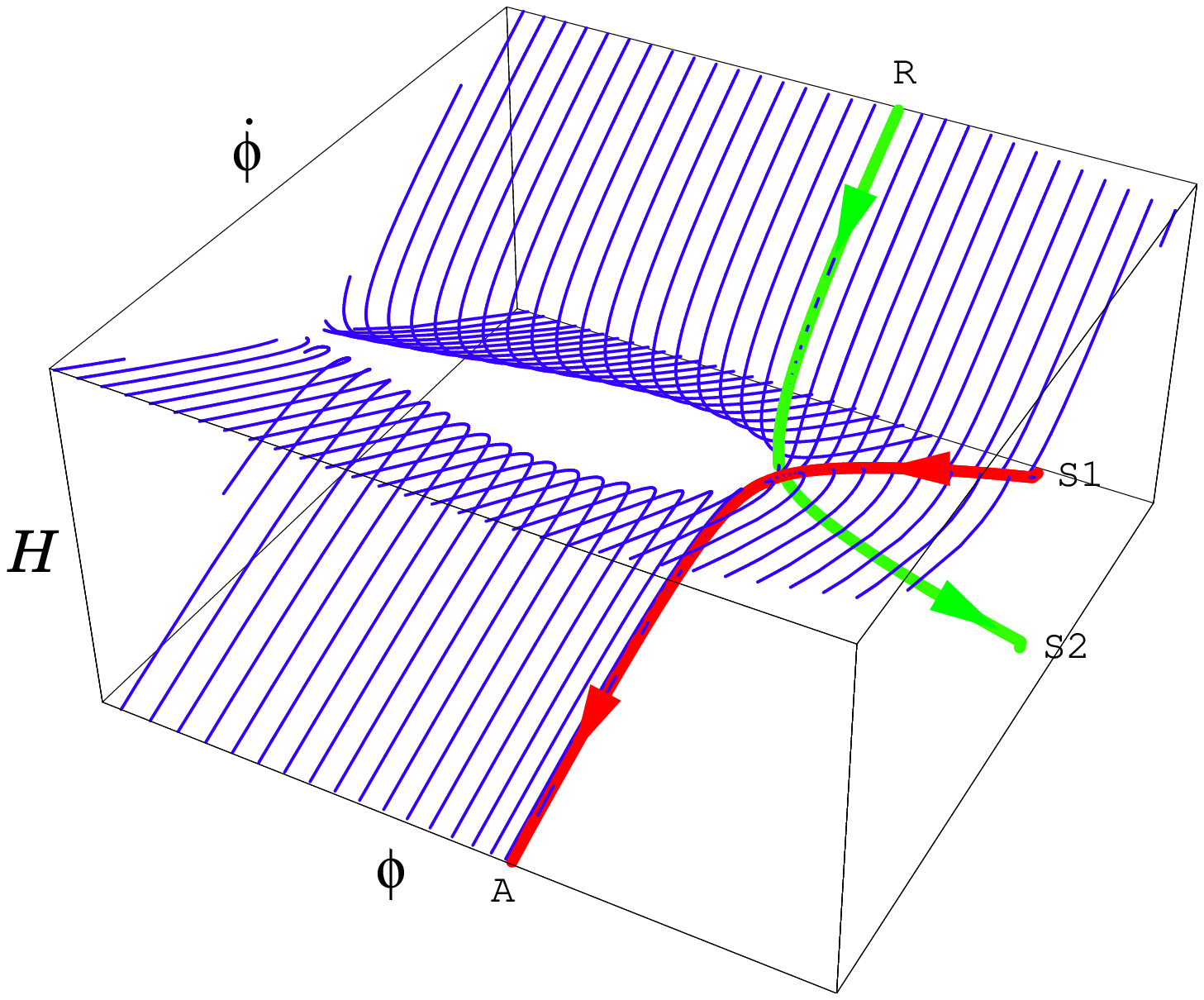,width=0.8\columnwidth}
\caption{The 3d phase portrait for the cyclic scenario. All
trajectories (blue lines) begin at $H > 0$ and end in a singularity at
$H<0$.}
\label{cycport}}

If one flips $\dot\phi \to -\dot \phi$ and $H \to -H$, which
corresponds to time-reversal, the red separatrix connecting points S1
and A becomes the green separatrix connecting points R and S2. In the lower
part of the figure (at negative $H$) this line corresponds to the
stage of deflation (exponential contraction of the universe, which is
a time-reversal of inflation).  These two separatrices divide all trajectories into three topologically disconnected parts: the trajectories to the right of the green separatrix, the trajectories between the green and the red separatrix and the trajectories to the left of the red separatrix.

One could think that the green separatrix separates inflationary trajectories from the trajectories that fall to the singularity without reaching the stage
of inflation. However, it is not so. As we already discussed in
Section \ref{negative}, the trajectories that reach the stage of
inflation are at a finite distance to the right away from the
green line connecting points R and S2 (i.e. at greater values of
$\phi$ and $\dot\phi$).

The $(\phi, \dot \phi)$ projection of the 
 phase portrait for the cyclic scenario is shown in
Fig. \ref{2dcycport}, also without the Poincar\'{e} mapping. An
interesting feature of the right panel of Fig. \ref{2dcycport} is the
apparent absence of any trajectories near the green line 
(the right separatrix at the right panel). This might
seem surprising because this line is a solution of the equations
of motion, so there must be other solutions nearby. The reason is that
the deflationary universe regime described by this line is a strong
repulsor, just opposite to the fact that the inflationary red line at
$H > 0$ (the right separatrix at the left panel) is a strong attractor. As a result, the density of
trajectories near the green line at $H <0$ is very small; that is why
they do not show up in Fig. \ref{2dcycport}. We discussed a similar
issue in Section \ref{negative}.

\FIGURE[!h]{ \epsfig{file=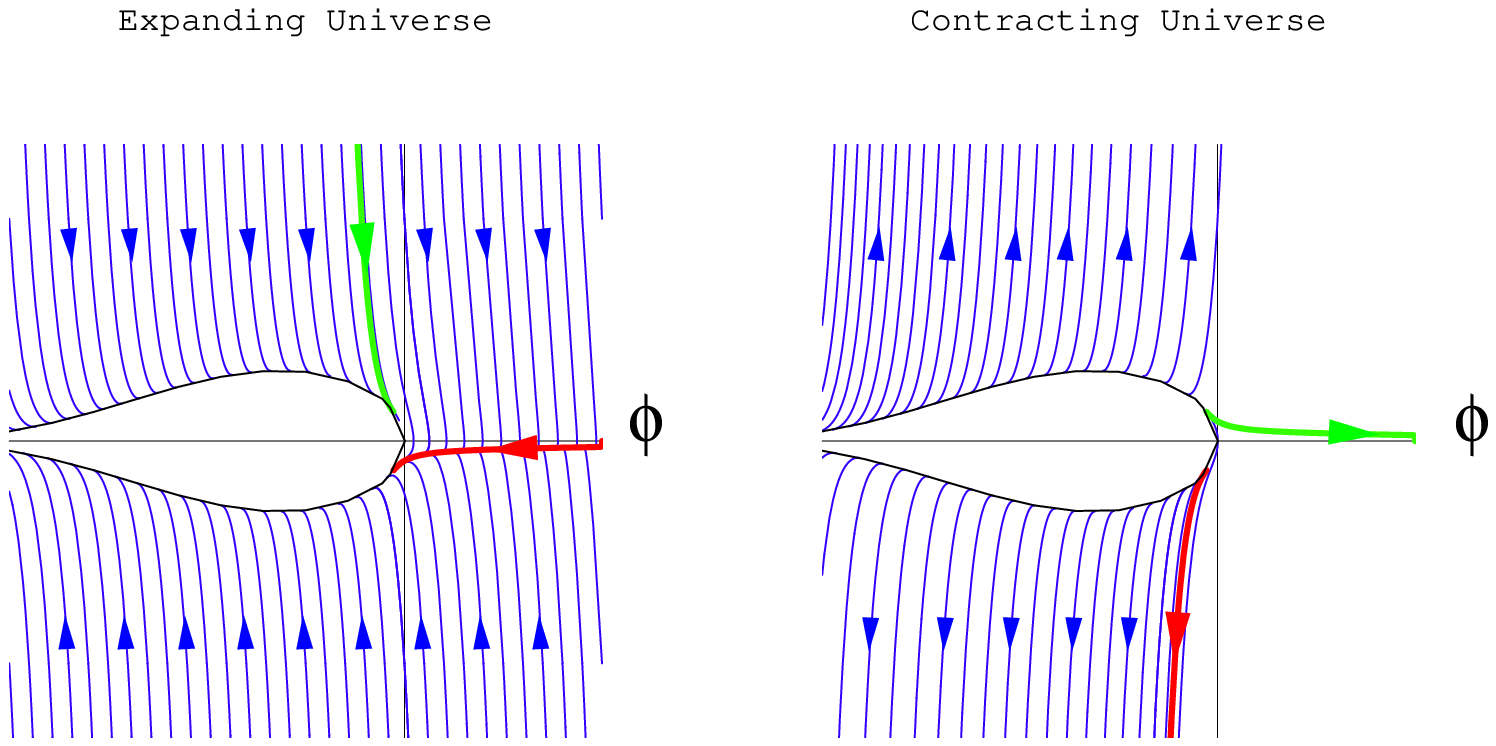,width=1.05 \columnwidth}
\caption{The 2d phase portrait for the cyclic scenario. All
trajectories begin at the bounding box of the left panel ($H>0$) and
end at the bounding box of the right panel ($H<0$). }
\label{2dcycport}}

As we see, all trajectories beginning at $H > 0$ end up in the
singularity at $H \to -\infty$. In the cyclic scenario it is assumed
that the universe goes through the singularity and re-appears
again. When it happens, all trajectories with $\phi<0$, $\dot\phi <0$
and $H<0$ the left lower part of the right panel in
Fig. \ref{2dcycport} suddenly reappear in the right upper corner of
the left panel of Fig. \ref{2dcycport}, describing the trajectories
starting at $\phi<0$, $\dot\phi >0$ and $H>0$. If one ignores particle
production at the singularity, the red separatrix on the right panel
becomes the green line at the left panel (time-reversal). As a result
of this flip, the field $\phi$, which previously was running down
along the red separatrix towards the singularity in Fig. \ref{cycport},
eventually returns exactly to the same place at $\phi >0$ where it was in the very
beginning of the process. However, it returns back not at the
stage of exponential expansion but at the stage of exponential
contraction, following the green separatrix in Fig. \ref{cycport}.

Exponential contraction is not a desirable regime. In order to 
reach the cyclic inflationary regime, some of the
trajectories to the left of the red separatrix after the singularity
should jump sufficiently far away to the right of the green
separatrix. As we already mentioned, Ref. \cite{Cyclic} assumes that
this jump may occur due to an increase in the energy of the scalar
field bouncing back from the singularity. This increase in energy is
supposed to happen due to particle production.  Only if this jump
is sufficiently large can these trajectories reach the red inflationary
separatrix going from S1 to A. Then inflation begins, the field rolls
to the minimum of $V(\phi)$ again, and everything repeats.

\subsection{\label{moving}Moving towards the minimum of $V(\phi)$}
 
To study the potential shown in Figure \ref{Cyclic} we will assume
that near the minimum it can be represented as ${m^2\over 2}(\phi^2
-\phi_0^2)$. At $\phi \gtrsim \phi_0$ we will take it to be flat with
$V \approx 10^{-120}$ and at $\phi < \phi_0$ we will take $V=0$. The
results of a numerical investigation for more complicated potentials
are very similar to the ones obtained for this simple model. However,
in this model one can study everything analytically using the results
obtained in Section \ref{switch}.  Indeed, we know how the field moves
at $\phi < -\phi_0$, when $V(\phi) = 0$, and we also know how it
behaves in the quadratic potential, when it moves from $-\phi_0$ to
$\phi_0$. The only thing that we need to do is to patch these two
regimes together.

At the initial stage the scalar field moves extremely slowly at $\phi
> \phi_0$ and the universe inflates. Once it reaches $\phi \approx
\phi_0$ it falls down, $V(\phi)$ becomes negative, and the universe
begins to contract. To describe this process one can use the theory
 developed in the first part of this paper. The contraction begins at $\phi=\phi_c$
(\ref{turning}).  The scalar field reaches $\phi = -\phi_0$ with
energy $\Delta\rho_-$ given by Eq. (\ref{engrowth}).

Subsequently, the field $\phi$ moves towards $\phi = -\infty$ and the
singularity develops in accordance with Eq. (\ref{1.7.25a}).  To
describe this motion one should take $t_0 = 1/\sqrt{3\Delta \rho_-}$ in
(\ref{1.7.25a}) and replace $\phi_0$ by $-\phi_0$:
\begin{eqnarray}\label{kinetdom2}
\phi + \phi_0 = \sqrt{2\over 3}\ \ln {\sqrt{3\Delta \rho_-}\, t }\ ,
\qquad {\dot\phi^2\over 2} = {1\over 3t^2}\ .
\end{eqnarray}
In this solution $\phi = -\phi_0$ at $t = t_0$.

Let us use this equation to find the value of the field $\phi$ at the
Planck time when the energy density becomes $1$ in Planck units and
one can no longer study this regime within the context of general
relativity. This happens at $t_p = 1/\sqrt 3$ in Planck
units. Therefore the scale factor of the universe $a \sim t^{1/3}$
decreases by a factor $\sim (\Delta \rho_-)^{1/6}$ from the beginning of
the process at $\phi = -\phi_0$ until the density becomes $O(1)$. The
scalar field $\phi$ at that time is given by
\begin{eqnarray}\label{kinetdom3a}
\phi_p = - \phi_0 + \sqrt{1\over 6}\ \ln {\Delta \rho_-}  \  .
\end{eqnarray}

Setting $a(t_p) = 1$ we can write our solution as
\begin{eqnarray}\label{kinetdom3}
\phi - \phi_p = \sqrt{2\over 3}\ \ln {\sqrt 3 t}\ , \qquad
{\dot\phi^2\over 2} = {1\over 3t^2}\ , \qquad a = 3^{1/6}\, t^{1/3} \
,
\end{eqnarray}
which in turn implies
\begin{eqnarray}\label{kinetdom33}
\dot\phi = {\sqrt 2\over a^3}, \qquad \phi - \phi_p = \sqrt{6 }\ \ln\,
a \ .
\end{eqnarray}

One can also represent our results in terms of the conformal time
$\tau$, where $dt = a d\tau$.  In this case $t = (2\tau\,
3^{-5/6})^{3/2}$, and
\begin{eqnarray}\label{kinetdom4}
\phi - \phi_p = \sqrt{3\over 2}\ \ln\, {2\, \tau\over \sqrt 3} \ .
\end{eqnarray}
The Planck time $t_p = 1/\sqrt 3$ corresponds to $\tau_p = \sqrt 3/2$.

The cyclic scenario requires that the universe bounce back from the
singularity and the field move back from $-\infty$ to
$\phi_0$. Depending on how much kinetic energy the field has at this
point three regimes are then possible:

\begin{enumerate}
\item ${\dot\phi^2\over 2} \le \Delta \rho_-$ at $\phi = -\phi_0$. This
is the regime that would be reached if the bounce were perfectly
symmetric (in which case ${\dot\phi^2\over 2} = \Delta \rho_-$). The
universe starts collapsing at $\phi \le \phi_c$. The field overshoots
the point $\phi = \phi_0$ and moves with ever growing speed towards
$\phi = +\infty$. There is a small bunch of trajectories such that the
scalar field evolves very slowly, the equation of state is $p =
-\rho$, and the universe {\it contracts} exponentially. Eventually,
however, the kinetic energy of the field $\phi$ dominates and the
collapse becomes power-law with $p = \rho$. 

This regime is represented by the trajectories to the left of the green separatrix in the upper part of the left panel in Fig. \ref{2dcycport}.

\item $\Delta \rho_- < {\dot\phi^2\over 2} < \Delta \rho_+$ at $\phi =
-\phi_0$. The universe starts collapsing at $\phi > \phi_c$. The field
does not have enough energy to reach the point $\phi = \phi_0$, so it
returns back to negative $\phi$, the field moves with ever growing
speed to $\phi = -\infty$, and a singularity develops.

This regime is represented by a small  bunch of  trajectories  to the right of the green separatrix in the upper part of the left panel in Fig. \ref{2dcycport}.

\item ${\dot\phi^2\over 2} \gtrsim \Delta \rho_+$ at $\phi =
-\phi_0$. The universe continues expanding and the field $\phi$
becomes greater than $\phi_0$. It continues growing and gradually
slows down. As a result, inflation begins. Then the field very slowly
decreases, falls into the minimum of $V(\phi)$, the universe collapses
and the field moves to $\phi = - \infty$. This is the regime required
by the cyclic scenario.

This regime is represented by   trajectories starting sufficiently far  from the green separatrix, to the right of it in the upper part of the left panel in Fig. \ref{2dcycport}.
\end{enumerate}

The last of these regimes requires additional explanation. Let us
remember how we derived the expression for $\Delta \rho_+$: We considered
the field $\phi$ slowly rolling from $\phi = \phi_0$ during the stage
of contraction and found that it arrived at the point $\phi = -\phi_0$
with kinetic energy $\Delta \rho_+$. If we reverse the time evolution of
the universe, we will see the scalar field rolling down from $\phi =
-\phi_0$ and arriving at the point $\phi= \phi_0$ with a nearly
vanishing speed {\it during the stage of expansion}. If the initial
kinetic energy of the field is greater than $\Delta \rho_+$, it reaches
the point $\phi= \phi_0$ with a non-vanishing speed and moves further
onto the plateau where the energy density of the field $\phi$ becomes
constant, and inflation begins.

As we have seen in Section \ref{switch}, the difference between
$\Delta \rho_+$ and $\Delta \rho_-$ is extremely small:
\begin{eqnarray}\label{diffen}
\delta \rho = \Delta \rho_+ - \Delta \rho_- = \pi \sqrt{3\Delta V}\,m\phi_0^2 \
.
\end{eqnarray}
Here $\Delta V$ has the meaning of the height of the effective
potential at $\phi > \phi_0$; in our case $\Delta V \sim
10^{-120}$. Thus one might expect that it pretty easy to jump from the
trajectory with energy $\Delta \rho_-$ to the desirable trajectory with
energy greater than $\Delta \rho_+$, as in case 3.

In reality, however, the required jump in kinetic energy becomes much
larger when one takes into account quantum effects. As the field
$\phi$ moves through the minimum from $-\phi_0$ to $\phi_0$ its mass
changes from $0$ to $m$ and back to $0$ again, all within a time
$O(m^{-1})$ (half of an oscillation), see 
Fig. \ref{Cyclic2}.
 This non-adiabatic change,
${\Delta m\over \Delta t} \sim m^2$, will lead to the production of
$\phi$-particles with energy density $O(m^4)$ \cite{KLSpreh}.
Therefore the field $\phi$ loses an amount of energy $O(m^4)$, which
makes it less likely to reach $\phi_0$ while the universe is still
expanding.\footnote{Note that the  production of $\phi$-particles during this very short
time interval appears in addition to the process of particle creation 
near the singularity discussed in Section \ref{singularity}.} Thus in order to realize the cyclic scenario the kinetic
energy density of the field $\phi$ at the point $-\phi_0$ must be
greater than $\Delta \rho_+$ by $O(m^4)$, which is much greater than
$\Delta V$. 

One may wonder where the field gets this boost in kinetic energy
from. Usually one would expect that the field after a bounce can only
lose energy due to particle production. However, in \cite{Cyclic} it
is assumed that it can actually gain energy as a result of particle
production during the brane collision (i.e. in the singularity). It is
not quite clear whether this can indeed happen, see
e.g. \cite{Rasanen:2001hf} where it is claimed that particles can be
created during the brane collision only if they have negative energy
density. We are not going to discuss this issue here. Instead of that,
we will follow the assumptions of \cite{Cyclic} and check what happens
to the scalar field $\phi$ if the universe after the bounce contains
some matter or radiation.

\subsection{\label{afterbounce}A scalar field with a vanishing
potential in the presence of radiation}

Let us consider the motion of the field $\phi$ from $-\infty$ to
$-\phi_0$ in the presence of radiation. The Friedmann equation
describing this process can be written as follows:
\begin{eqnarray}\label{friedrad}
\left({\dot a \over a}\right)^2= {1\over 3}\left({\dot\phi_i^2\over
2}\ {a^6_i\over a^6} + \rho_i^r\ {a^4_i\over a^4}\right) \ .
\end{eqnarray}
Here $\dot\phi_i$ is the velocity of the field at some moment $t_i$,
$a_i$ is the scale factor of the universe at that moment, and
$\rho_i^r$ is the density of radiation at that time. This equation
reflects the fact that the kinetic energy of the field decreases as
$a^{-6}$ and radiation energy decreases as $a^{-4}$ during the
expansion of the universe.\footnote{Here we are considering processes
at sub-Planckian energies where the usual Friedmann cosmology is
supposed to be valid.}

It is convenient to write this equation in terms of the conformal time
$\tau$, where $dt = a d\tau$:
\begin{eqnarray}\label{frradconf}
(a')^2 = {A^2\over  a^2} + B \ ,
\end{eqnarray}
where $a' = {da\over d\tau} = a\,\dot a $, $A^2={\dot\phi_i^2 a^6_i/
6}$ and $B={\rho_i^r a^4_i/ 3}$.

 Taking $a(0)=0$ (at the singularity), the solution of this equation is
\begin{eqnarray}\label{scaleconf}
a^2 = {2A\tau} + B\tau^2 \ .
\end{eqnarray}

For definiteness, we will normalize our solution at the time $t_i=
t_p$, when $\dot\phi^2/2 = 1$ and $a_i = 1$. Then $A^2={1/ 3}$,
$B={\rho_p^r / 3}$, and
\begin{eqnarray}\label{scaleconf2}
a^2 = {{2\over \sqrt 3} \tau} + {\rho_p^r \over  3}\tau^2 \ .
\end{eqnarray}

Then, using equation $\phi' = \dot\phi_i a_i^3/a^2 = \sqrt 6 A/a^2$,
one finds
\begin{eqnarray}\label{scaleconf3}
\phi -\tilde{\phi}_p = \sqrt{3\over 2}\, \ln {2 \tau\over \sqrt 3
(1+{\rho_p^r\over 2\sqrt 3}\tau)} +C_r =\sqrt{6}\, \ln {a\over
1+{\rho_p^r\over 2\sqrt 3}\tau } + C_r\ .
\end{eqnarray}
Here $\tilde{\phi}_p$ is the value of the scalar field at the time when
$\dot\phi^2/2 = 1$ after the bounce. The constant of integration $C_r$
is supposed to vanish in the absence of radiation, i.e. for $\rho_p^r
= 0$. In this case $\tilde{\phi}_p= \phi_p$, and our solution
(\ref{scaleconf3}) coincides with the solution presented in
Eq. (\ref{kinetdom4}). This means that in the absence of radiation the
field $\phi$ elastically bounces from the singularity, in accordance
with \cite{Seiberg}.

One can find the constant $C_r$ for any given $\rho_p^r$ from the
condition that $\phi = \tilde{\phi}_p$ at $\dot\phi^2/2 = 1$ and $a = 1$. In
particular, for $\rho_p^r \ll 1$ one has $C_r \approx \rho_p^r {\sqrt
3\over 2\sqrt 2}$.

Eq. (\ref{scaleconf3}) implies that at $ \rho_p^r \tau > 2\sqrt 3$ the
field stops moving. Therefore we will assume that $ \rho_p^r \tau \ll
1$ at $\phi < -\phi_0$.  This leads to a strong constraint on
$\rho_p^r$:
\begin{eqnarray}\label{scaleconf3c}
\rho_p^r \lesssim (\Delta \rho_+)^{1/3}\  .
\end{eqnarray}
If one takes, for definiteness, $\phi_0 \sim 0.1 M_p$, $m^2\phi_0^2
\sim 10^{-20}$, as in the original version of the ekpyrotic scenario
\cite{KOST}, one finds that the cyclic scenario with these parameters
cannot work unless the energy density of radiation at the Planck time
is less than $10^{-6}$ in Planck units. In general the density of
gravitationally produced particles is $\sim H^4$, which is $O(1)$ at
the Planck time, so it is not clear how particle production could be
so strongly suppressed.

Suppose, however, that for whatever reason one can indeed have $
\rho_p^r \ll (\Delta \rho_+)^{1/3}$.  In this case $\rho_p^r \tau \ll
1$ and Eq. (\ref{scaleconf3}) can be represented in the following
form:
\begin{eqnarray}\label{scaleconf3a}
\phi -\tilde{\phi}_p = \sqrt{6}\, \ln a -{\rho_p^r\over \sqrt 2}\, (\tau-
{\sqrt 3 \over 2} )\ .
\end{eqnarray}
With our normalization of $a$ one has
\begin{eqnarray}\label{scaleconf5}
{1\over 2} \dot\phi^2   =  a^{-6}  \ ,
\end{eqnarray}
As we already discussed, if we want the field to move to $\phi >
\phi_0$ during the stage of expansion of the universe, its kinetic
energy $\dot\phi^2/2$ must be greater than $\Delta \rho_+$ at $\phi =
-\phi_0$. If we assume that the field has sub-Planckian energy as it
moves through the minimum, i.e. that $\Delta \rho_+ \ll 1$, then
\begin{eqnarray}\label{scaleconf3a1}
  \tilde{\phi}_p > -\phi_0 + {1\over \sqrt 6}\, \ln \Delta \rho_+ +{\rho_p^r \sqrt
3\over 2 \sqrt 2}\, (\Delta \rho_+)^{-1/3} \ .
\end{eqnarray}
Comparison with Eq. (\ref{kinetdom3a}) gives the following condition:
\begin{eqnarray}\label{scaleconf3a2}
\tilde{\phi}_p -\phi_p > {\rho_p^r \sqrt 3\over 2 \sqrt 2}\, (\Delta
\rho_+)^{-1/3} + {1\over \sqrt 6}\, \ln {\Delta \rho_+\over \Delta \rho_-} \ .
\end{eqnarray}
In general, it could happen that after bouncing from the singularity
the field $\phi$ appears at the Planck density at $\tilde{\phi}_p \not
=\phi_p$, so that $\tilde{\phi}_p -\phi_p = O(\rho_p^r)$
\cite{Cyclic}. However, our investigation shows that the cyclic
scenario with $ \Delta \rho_+ \ll 1$ could work only if $\tilde{\phi}_p -\phi_p
\gg \rho_p^r$.

This means that the cyclic scenario can work only if a very small
amount of radiation can produce a major change in the state of the
field $\phi$ at the Planck time: $\tilde{\phi}_p -\phi_p \gtrsim \rho_p^r \,
(\Delta \rho_+)^{-1/3}$. Second, the amount of radiation at the Planck
time must be very small, $\rho_p^r \lesssim (\Delta \rho_+)^{1/3}$. This
may be a real problem if, as we expect, quantum effects at Planckian
densities create particles with density $\rho_p^r = O(1)$.

These problems are less serious in models with $ \Delta \rho_+ \geq 1$,
i.e. if the field $\phi$ acquires super-Planckian energy even before
it reaches the plateau at $\phi < -\phi_0$. Such models are suspect
because the usual 4D approach based on general relativity becomes
unreliable at super-Planckian densities. It appears that such models
are necessary for the cyclic model, however, and in at least one of
their papers the authors of \cite{Cyclic} invoke such a model. We
therefore consider such potentials here.

\subsection{\label{superplanck}Super-Planckian potentials for the
cyclic scenario}

Let us now consider a potential proposed by the authors of the cyclic
scenario \cite{Cyclic}:
\begin{eqnarray}\label{realpot}
V(\phi) =  V_0~(1-e^{-c\phi})~ F(\phi) \ .
\end{eqnarray}
In the particular example studied in the last paper of
Ref. \cite{Cyclic} one has $F(\phi) = e^{-e^{-\gamma \phi}}$, $V_0 =
10^{-120}$, $c = 10$, $\gamma \approx 1/8$. This potential is shown in
Fig.  \ref{Cyclicfig}. This potential has the same structure
as the potential shown in the Fig. 12, but the scales and the position of the minimum are determined by the
parameters given in \cite{Cyclic}. 
  At $\phi =0$ this potential vanishes. It
approaches its asymptotic value $V_0 = 10^{-120}$ at $\phi \gtrsim
1$. Inflation in this scenario is possible at $\phi \gtrsim 1$. At
$\phi \gtrsim 15$ one has $V^{3/2}\gtrsim V'$ and the universe enters
the process of eternal inflation \cite{VilEt,Eternal}. The potential
has a minimum at $\phi \approx -36$; the value of the potential in
this minimum is $V_{\rm min} \approx -3$.

\FIGURE[!h]{\epsfig{file=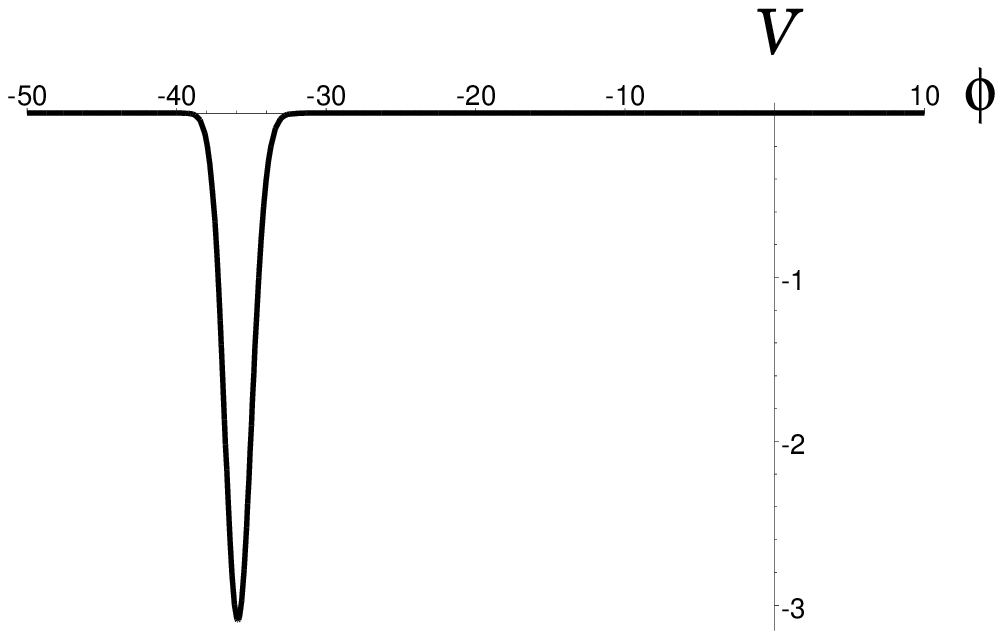,width=.6\columnwidth}
\caption{An example of cyclic scenario potential used in Ref. \cite{Cyclic}.}
\label{Cyclicfig}}

Let us try to understand the origin of the parameters $c = 10$,
$\gamma \approx 1/8$ used in \cite{Cyclic}.  According to
\cite{Khoury:2001zk}, the amplitude of density perturbations in this
scenario in the limit $c\gg 1$ can be estimated as
\begin{eqnarray}\label{denspert}
{\delta\rho\over \rho} \sim  10^{-5} \sqrt{-V_j}~\xi^4 \ ,
\end{eqnarray}
where $V_j$ is approximately equal to the value of the potential in
its minimum $V_{\rm min}$ and $\xi$ is the efficiency with which
radiation is produced at the singularity; it is assumed that $\xi \ll
1$. This suggests that in order to be consistent with observational
data (${\delta\rho\over \rho} \sim 10^{-4}$) one should have $-V_j \gg
1$. This means one must rely on calculations using the equations of
general relativity at $|V(\phi)| \gg 1$.

The authors of \cite{Khoury:2001zk} have warned the readers that their
results are very preliminary and many authors do not agree with their
derivation of the amplitude of density perturbations
\cite{Lyth:2001pf}.  Therefore it may happen that the correct equation
for perturbations in the cyclic scenario as well as the expression for
$V(\phi)$ will be quite different. Here we will simply try to
understand the values of the parameters used in \cite{Cyclic} and
check the consequences of the potential they suggested.

The spectrum of density perturbations obtained in \cite{Khoury:2001zk}
is not blue, as in \cite{KOST}, but red, like in the pyrotechnic
scenario \cite{pyrotech} and in the simplest versions of chaotic
inflation. The spectral index is $n \approx 1-4/c^2$. Observational
data suggest that $n = 0.93 \pm 0.1$, which implies that $c \gtrsim
5$. If one takes $c \gg 5$, and $V_J >1$, one finds that the curvature
of the effective potential in its minimum becomes much greater than 1.

Once one takes $ V \sim -3$ in the minimum of the potential with $c =
10$ \cite{Cyclic}, the parameter $\gamma$ can be determined
numerically: $\gamma = 0.1226$.
It would be  hard to provide explanation of the numerical value
of this parameter. Meanwhile if one takes $\gamma = 1/8=
0.125$, one finds $ V \sim -3\times 10^{-3}$ in the minimum of the
potential. This would reduce ${\delta\rho\over \rho}$ by a factor of
30. Thus, in order to have density perturbations with a correct
magnitude one should fine-tune the value of $\gamma = 0.1226$ with
accuracy better than 1\%.

\FIGURE[!h]{\epsfig{file=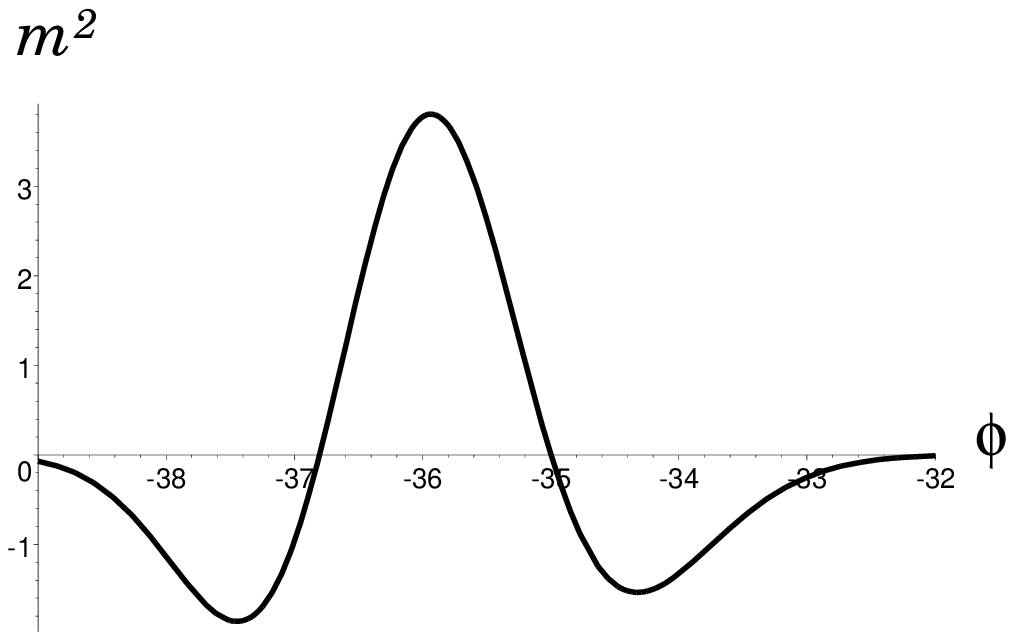,width=.6\columnwidth}
\caption{Effective mass squared $m^2 = V''$ of the scalar field in the
vicinity of the minimum of $V(\phi)$ in the cyclic scenario.}
\label{Cyclic2}}

Fig. \ref{Cyclic2} shows the effective mass of the field $\phi$. As we
see, $|m^2| = |V''| \gtrsim 1$ in the vicinity of the minimum of the
effective potential.   A numerical
investigation of the motion of the field moving from $\phi >0$
in a theory with this potential shows that its kinetic energy at the
moment when $\phi$ reaches the minimum of the effective potential is
$O(10^2)$. When the field approaches $\phi \sim -39$, where the
effective potential becomes flat, the kinetic energy of the field
$\phi$ becomes $\sim 10^6$, i.e. a million times greater than the
Planck density.

Even if we continue to trust our calculations in such a regime, there
are still problems. First of all, there is a distance $\Delta\phi >
30$ from the point  $\phi \approx -30$ where the field emerges from
the deep minimum of its effective potential to the region $\phi >1$,
where inflation in this theory may begin. Let us assume that the
kinetic energy of the field is smaller than the Planck energy at $\phi
\sim -30$, since otherwise we just cannot trust our analysis at
all. This assumption is in accordance with \cite{Cyclic}. Indeed,
according to the estimates made in \cite{Cyclic},
$\tilde{\phi}_p-\phi_p \approx \ln{H_5(out)\over H_5(in)} < {1\over 2}
\ln {4\over 3} <1 $. In this model $\phi_p \approx -34$, so indeed one
expects $\tilde{\phi}_p < -33$.

As we discussed in Section \ref{afterbounce}, we expect that
gravitational particle production will create particles with density
$O(1)$ at the Planck time. Independently of gravitational production,
however, there should be production of $\phi$ particles with density
$O(m^4)$ due to the nonadiabatic change of the effective mass of the
field moving from $\phi = -39$ to $\phi > -32$, see
Fig. \ref{Cyclic2}. In this model $O(m^4) \gtrsim O(1)$. Thus, when
the field reaches the relatively flat region at $\phi > -32$, its
motion produces ultra-relativistic particles $\phi$ with
super-Planckian energy density. These particles, just like usual
radiation, immediately freeze the motion of the field $\phi$. One can
show that in this scenario the field $\phi$ can reach the inflationary
regime at $\phi > 0$ (which is necessary for the consistency of the
cyclic scenario) only if at $\phi \sim -32$ (i.e. at the flat part of
the potential) the kinetic energy density of the field $\phi$ is 12
orders of magnitude greater than the (Planckian) energy density of the
produced particles. The effective 4D description in terms of the
scalar field $\phi$ and its effective potential $V(\phi)$ is
inapplicable for the description of such processes.

This problem is not unresolvable. For example, one may consider
effects related to non-relativistic particles produced at the
singularity. These particles contribute to the equation of motion for
the field $\phi$ by effectively increasing its potential energy
density \cite{Cyclic}. They may push the field towards positive values
of the field $\phi$ despite the effects described above. However, this
would add an additional epicycle to a scenario that is already quite
speculative.  Indeed, one would need to produce a sufficiently large
number of such particles and make sure that massive particles decouple
from the scalar field at the present epoch. The last condition is
necessary to avoid a rapid change of the coupling constants related to
the brane separation described by the field $\phi$.

One may try to improve the situation by altering the shape of the
potential. First of all, the original argument of \cite{Cyclic} was
that the function $F(\phi)$ appears because at small values of the
string coupling $g_s$ nonperturbative effects should be suppressed by
a factor $e^{-1/g_s}$ or $e^{-1/g^2_s}$, or perhaps by
$e^{-8\pi^2/g^2_s}$. In the case of type IIA (or heterotic) string
theory in d=10 the string coupling is $g_s = e^{-\phi}$
\cite{Seiberg}. Thus one could expect the suppression function to be
one of the three proposed types: $F(\phi) \sim e^{-e^{-\gamma \phi}}$,
$F(\phi) \sim e^{-e^{-2\gamma \phi}}$, or $F(\phi) \sim e^{-8\pi^2
e^{-\gamma \phi}}$, with $\gamma = 1$ rather than with $\gamma=0.1226$.

It is possible to have $V_{\rm min} = -3$, as in \cite{Cyclic}, for
$\gamma = 1$, but only if one takes $c = 81.56$. The value of $c$ must
be fine-tuned: a change in $c$ of 1\% results in a change of $V_{\rm
min}$ by two orders of magnitude. In accordance with
\cite{Khoury:2001zk}, this would lead to an order of magnitude change
in the amplitude of density perturbations.

With these parameters, however, the curvature of the effective
potential in its minimum becomes two orders of magnitude greater than
the Planck mass squared, so all calculations in such models in the
context of the effective 4D theory are unreliable. In potentials with
$F(\phi) \sim e^{-e^{-2\gamma \phi}}$ or $F(\phi) \sim e^{-8\pi^2
e^{-\gamma \phi}}$ the curvature in the minimum with $|V(\phi)|
\gtrsim O(1)$ becomes much greater still.

\subsection{\label{bicycling} Bicycling scenario}

Various modifications to the cyclic scenario are possible. For
example, instead of the asymmetric potential shown in
Figs. \ref{Cyclic} and \ref{Cyclicfig}, one may consider a symmetric
potential, as in Fig. \ref{Cyclic3}.

\FIGURE[!h]{\epsfig{file=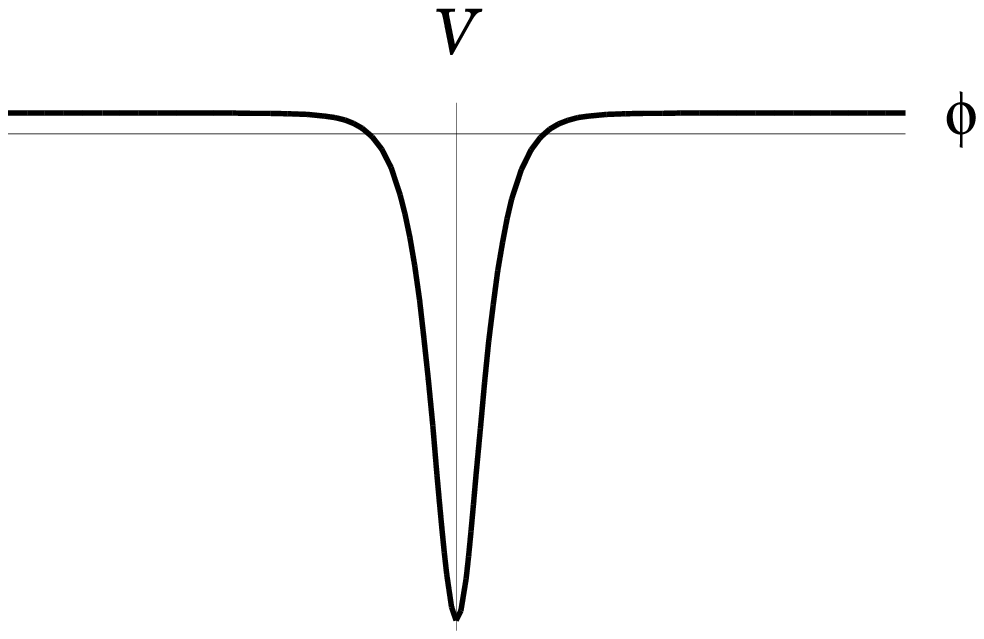,width=.6\columnwidth}
\caption{Symmetric scalar field potential in the new cyclic
scenario. At large values of $|\phi|$ one has $V(\phi) \approx V_0
\sim 10^{-120}$ and there is a minimum at $\phi = 0$.}
\label{Cyclic3}}

In the beginning, the scalar field is large and positive and it slowly
moves towards the minimum. When it falls to the minimum the universe
begins to contract and the field is rapidly accelerated towards the
singularity at $\phi = -\infty$. As we already mentioned, the
structure of the singularity is not sensitive to the existence of the
potential, especially if it is as small as $V_0 \sim
10^{-120}$. Suppose in the vicinity of its minimum the potential is
approximately quadratic, $V(\phi) \approx m^2(\phi^2-\phi_0^2)/2$.  If
$\phi_0 \leq 1$ and $m \ll 1$, then according to
Eq. (\ref{kinetdom3a}) the kinetic energy of the field $\phi$ reaches
the Planck value at
\begin{eqnarray}\label{kinetdom3anew}
\phi_p = - \phi_0 + \sqrt{1\over 6}\ \ln {\Delta \rho_-} \approx
\sqrt{2\over 3}\ \ln {(m \phi_0^2)} \ .
\end{eqnarray}
For definiteness, suppose that $m \sim \sqrt{V_0} \sim 10^{-60}$, and
$\phi_0 = O(1)$. Then we would not even know that such a minimum
exists (the field would not move there) until the energy density of
matter dropped below its present density $10^{-120}$. In this case the
kinetic energy of the field moving towards $\phi = -\infty$ would
reach the Planck value at $\phi_p \sim -112$. At that time the scale
factor of the universe would decrease by a factor of $\Delta
\rho_-^{1/6} \sim 10^{-20}$.

Now let us assume, as in \cite{Cyclic}, that the field $\phi$ bounces
from the singularity and moves back. Its energy density drops down to
the Planck energy density at $\tilde{\phi}_p \approx \phi_p \sim
10^2$. During its subsequent evolution the kinetic energy of the field
rapidly drops down because of radiation. Even if the density of
radiation at the time when $\phi = \tilde{\phi}_p$ were as small as
$10^{-39}$, it would eventually begin to dominate because its relative
contribution grows as $a^2$, i.e. up to $10^{40}$ times before it
reaches $-\phi_0$.

Therefore the field $\phi$ freezes at large negative $\phi$. At this
stage the energy density is dominated by particles produced near the
singularity and density perturbations prepared during the previous
cycle lead to structure formation. Then the universe cools down while
the field is still large and negative and the late-time stage of
inflation begins. During this stage the field slowly slides towards
the minimum of the effective potential and then rolls towards the
singularity at $\phi \to \infty$. When it bounces from the
singularity, a new stage of inflation begins. The universe in this
scenario enters a cyclic regime with twice as many cycles as in the
original cyclic scenario of Ref. \cite{Cyclic}. One may call it the
{\it bicycling scenario}.

An advantage of this scenario is that it may work even if a lot of
radiation is produced at the singularity and the field $\phi$ rapidly
loses its kinetic energy. However, if in order to have density
perturbations of a sufficiently large magnitude one needs to have a
potential with a super-Planckian depth $V(\phi) <-1$, as in
\cite{Khoury:2001zk,Cyclic}, then this scenario has the same problem
as the scenario considered in the previous section. The kinetic energy
of the field $\phi$ becomes greater than the Planck density as soon as
it rolls to the minimum of $V(\phi)$. It becomes even much greater
when the field rolls out of the minimum, and the 4D description fails.

\subsection{\label{cyclicinflation}Cycles with inflationary density
perturbations}

As we see, one of the main difficulties of the cyclic scenario is
related to the non-inflationary mechanism of generation of density
perturbations.  It requires a very specific and fine-tuned potential,
see \cite{pyrotech} and discussion above. According to
\cite{Khoury:2001zk}, this potential must have a super-Planckian
depth, so one cannot study the corresponding processes by traditional
methods. Moreover, the very existence of this mechanism of generation
of density perturbations remains controversial \cite{Lyth:2001pf}.

This problem  can  be avoided  if we
consider  a potential that grows at large $|\phi|$, such as the one shown in Fig. \ref{Cyclic5}. The
field begins to move from large positive $\phi$, falls to the minimum
of $V(\phi)$, and moves with ever growing speed to $-\phi$. If, for example, the
potential grows like $\phi^n$ at a sufficiently large negative $\phi$,
it does not affect the motion of the field $\phi$ towards the
singularity. However, when the field $\phi$ bounces back, it
immediately loses its velocity due to the impact of radiation created at
the singularity. Therefore it slows down {\it and enters a stage of
inflation}. At this stage all good and bad memory about the previous
life of the universe and processes at the singularity are erased and
new density fluctuations are produced. All particles produced at the
singularity become diluted, but new ones are produced at the end of inflation
due to gravitational effects \cite{QuintInfl} or by the mechanism of
instant preheating \cite{Instant,NO}. These new particles constitute the
matter contents of the observable universe.

\FIGURE[!h]{\epsfig{file=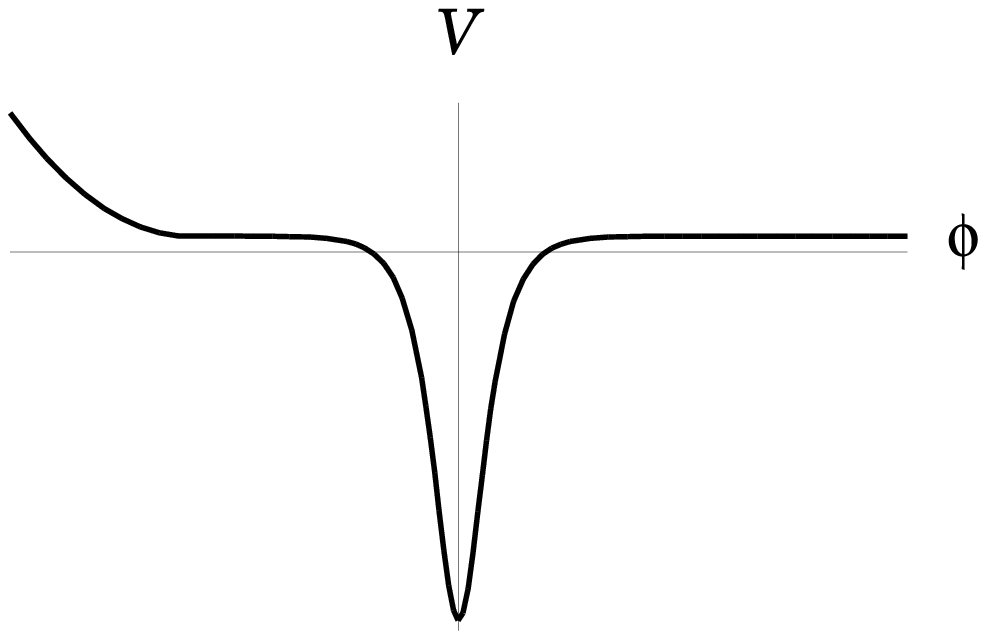,width=.6\columnwidth}
\caption{Scalar field potential in the cyclic scenario incorporating a
stage of chaotic inflation. Inflationary perturbations are generated and the large-scale structure of the universe is produced  at $\phi < 0$. }
\label{Cyclic5}}

Gradually the density of ordinary matter decreases, and the energy
density of the universe becomes determined by $V(\phi) \approx
V_0$. The universe enters a stage of low energy inflation (quintessence), which may
result in a regime of self-reproduction if $V(\phi)$ is flat
enough. In those exponentially large domains of the universe where the
field eventually falls down to the minimum of $V(\phi)$, it continues
rolling to $\phi = \infty$, bounces back after the singularity, slows
down due to radiation, experiences low-energy inflation, and rolls
down to the minimum of $V(\phi)$ again.

In this model of the oscillating universe one can have large scale
structure formation due to inflationary perturbations without any need
to rely on controversial assumptions about the behavior of
perturbations passing through the singularity. Also, one no longer
needs to have potentials with $|V(\phi)| > 1$.  However, in this model
inflationary perturbations are generated only every second time after
the universe passes the singularity (at $\phi <0$, but not at $\phi
<0$). The model can be made even better by making the potential rise
both at $\phi \to \infty$ and at $\phi \to -\infty$, see
Fig. \ref{Cyclic6}. In this case the stage of high-energy inflation
and large-scale structure formation occurs each time after the
universe goes through the singularity.

\FIGURE[!h]{\epsfig{file=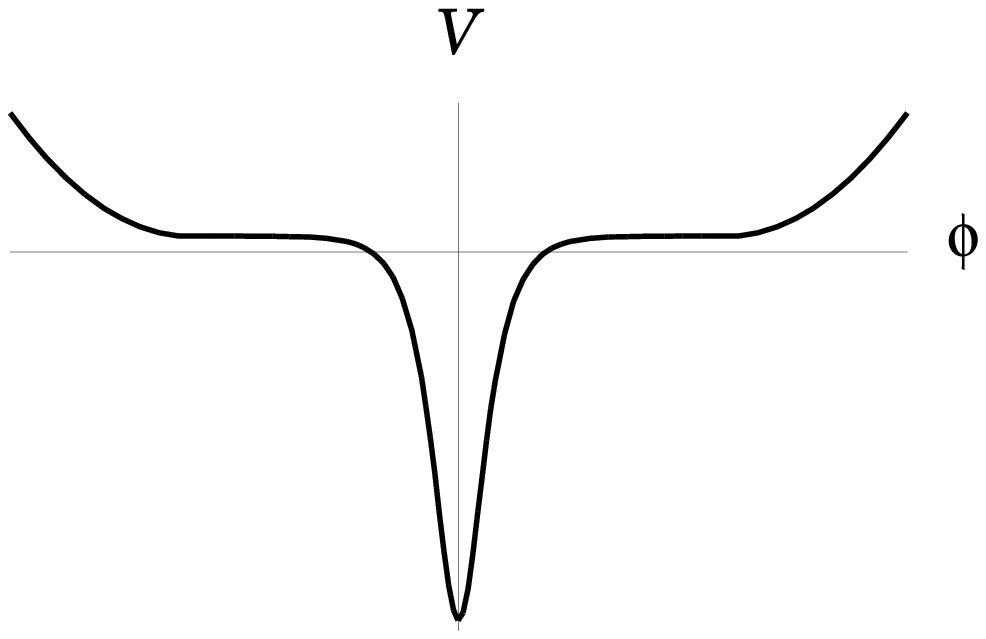,width=.6\columnwidth}
\caption{Scalar field potential in the cyclic scenario incorporating a
stage of chaotic inflation. Inflationary perturbations are generated and the large-scale structure of the universe is produced both at $\phi < 0$ and at $\phi > 0$.}
\label{Cyclic6}}

Thus we see that it is possible to propose a scenario describing an
oscillating inflationary universe without making any assumptions about
the behavior of non-inflationary perturbations near the
singularity. Another important advantage of this scenario is that
inflationary cycles may begin in a universe with initial size as small
as $O(1)$ in units of the Planck length, just as in the standard
chaotic scenario \cite{Chaot}. Still, in many other respects this
scenario is almost as complicated as the cyclic scenario of
Ref. \cite{Cyclic}. The theory of reheating of the universe in this
model, just as in \cite{Cyclic}, is rather unconventional. Gravitational
particle production, which is the only source of matter in this scenario,
 may dramatically overproduce
gravitinos and moduli fields \cite{QuintInfl,NO}. To avoid this
problem one would need to use the mechanism of instant preheating
\cite{Instant,NO}. In order to combine the stage of chaotic inflation
and the stage of low-scale inflation (quintessence) the potential must
be rather complicated. To avoid this complication one may need to
consider two-field models of the type of hybrid inflation.
 
\FIGURE[!h]{\epsfig{file=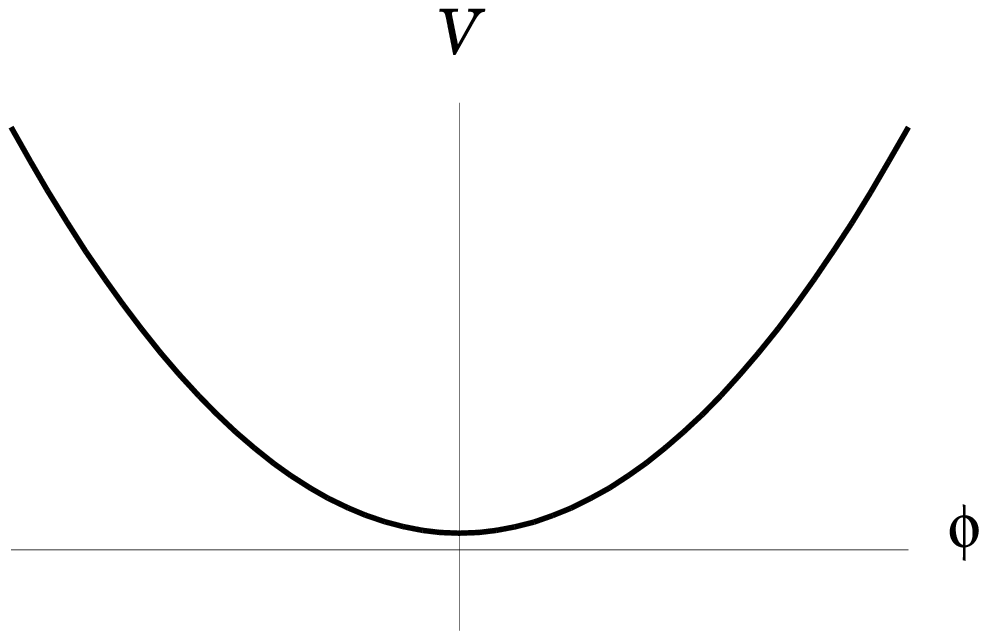,width=.6\columnwidth}
\caption{The scalar field potential that appears after the step-by-step simplification of the cyclic scenario.}
\label{Cyclic7}}

The main problem of this model is that one still must assume that
somehow the universe can go through the singularity. But now this
assumption is no longer required for the success of the scenario since
the large scale structure of the universe in this scenario does not
depend on processes near the singularity. This allows us to remove the
remaining epicycles of this model. Indeed, the main source of all the
problems in this model is the existence of the minimum of the
effective potential with $V(\phi)<0$. Once one cuts this minimum off,
the potential becomes extremely simple, see Fig. \ref{Cyclic7}, and
all problems mentioned above disappear. In particular, one may use the
simplest harmonic oscillator potential ${m^2\over 2}\phi^2 +V_0$ 
with $V_0 \sim 10^{-120}$
considered in the beginning of our paper. This theory describes an
eternally self-reproducing chaotic inflationary universe, as well as
the late stage of accelerated expansion (inflation) of the universe
driven by the vacuum energy $V_0>0$.

\section{\label{Conclusions}Conclusions}

The main goal of our work was to perform a general investigation of
scalar field cosmology in theories with negative potentials. We have
found that the phase portraits of such theories in the 3d space
$(\phi,\dot\phi,H)$ have different geometry  as compared with phase
portraits in theories with $V(\phi) \geq 0$. In theories with $V(\phi)
>0$ the phase portraits  for flat universes are divided into two
disconnected parts describing expanding and contracting universes ($H
>0$ and $H<0$). Meanwhile in theories with $V(\phi) <0$ these two
parts become connected. The trajectories moving towards $V(\phi) <0$
simultaneously move from the parts of the phase portrait with $H >0$
towards the parts with $H<0$. Once the universe begins to contract, it
never returns to the stage of expansion until it reaches the
singularity.

This does not mean that theories with negative potentials should be
banned from consideration. In some cases the scalar field may be
trapped in a metastable minimum, or it may roll towards $V(\phi)<0$
extremely slowly. However, it is quite interesting that with an
account taken of general relativity potentials that have  minima at
$V(\phi)<0$ can be as dangerous as potentials unbounded from below.

A general feature of all trajectories bringing the universe towards
the singularity is that in all theories with power-law potentials the
kinetic energy $\dot\phi^2/2$ becomes much greater than $V(\phi)$ near
the singularity. This means that the description of the singularity is
nearly model-independent, at least at the classical level. In
particular, the equation of state of the universe approaching the
singularity typically is $p = \rho$.

However, this conclusion can be altered with an account taken of
quantum effects, including particle production near the
singularity. Typically particle production near the singularity is so
efficient that it turns off the regime $p = \rho$ when a contracting
universe approaches the Planck density. The effects related to
particle production are especially significant in an expanding
universe as they tend to completely eliminate the stage with $p =
\rho$.

In addition to the general study of cosmology with negative
potentials, we performed an investigation of a possibility that our
universe may experience repeated cycles of inflation and contraction
\cite{Cyclic}. For a complete study of this scenario one would need to
resolve the singularity problem, as well as several other problems
discussed in \cite{pyrotech,KKLT,Lyth:2001pf,Rasanen:2001hf}. In
addition, as we show in this paper, the parameters of the effective
potentials used in the cyclic scenario must be fine-tuned with
accuracy better than 1\%. This scenario, as proposed in \cite{Cyclic},
requires investigation of an effective potential $V(\phi)$ of a
super-Planckian depth, $|V(\phi)|>1$, and of a scalar field with mass
greater than the Planck mass. Even if all of these problems could be
resolved in the context of a more general approach, the existence of a
cyclic regime in the model of Ref. \cite{Cyclic} would require
additional assumptions.  We have shown that ultrarelativistic
particles produced near the singularity, as well as scalar particles
created when the field falls down to the minimum of the effective
potential, tend to halt the motion of the classical field $\phi$,
which prevents inflationary cycles from occurring. One way to address
this problem is to study quantum creation of supermassive particles
with specific interactions with the scalar field.  However, this would
add new ``epicycles'' to a scenario that is already very complicated.

We proposed several modifications to the cyclic scenario of
Ref. \cite{Cyclic} that could make it more realistic and less
dependent on the unsolved singularity problem. In particular, if one
assumes that the potential $V(\phi)$ slowly grows at large $|\phi|$
then the universe may still enter a regime of eternal oscillations,
but the singularity will be separated from the stage of large scale
structure formation by a stage of chaotic inflation. This scenario
allows us to combine attractive features of the oscillating universe
model \cite{Tolman-1931} - \cite{LindeReview} and chaotic inflation
\cite{Chaot}. An important advantage of this model is that it does not
need to rely on the controversial theory of density perturbations
passing through the cosmological singularity.

But even this model remains very complicated. Fortunately, it allows
for one final simplification that resolves all of its remaining
problems.  If one removes the minimum of the potential at $V(\phi)
<0$, one returns to the usual scenario of chaotic inflation.  It
describes an eternally self-reproducing inflationary universe, as well
as the present stage of accelerated expansion.

\section*{Acknowledgements}
The authors are grateful to R. Brandenberger, R. Kallosh,
 A. Maroto, S. Rasanen, A. Starobinsky, and N. Turok for many interesting discussions. 
The work by G.F., A.F. and L.K. was supported by
PREA of  Ontario, NSERC and  CIAR.
 The work by A.L. was supported
by NSF grant PHY-9870115, by the Templeton Foundation grant
No. 938-COS273.  G.F., L.K. and A.L. were also supported
  by NATO Linkage Grant 975389.

\end{document}